\documentclass[prd,aps,twocolumn,nofootinbib]{revtex4-1}

\usepackage{graphicx}

\usepackage{hyperref}

\begin{document}
\title{Twist-3 light-cone distribution amplitudes of the scalar mesons within the QCD sum rules and their application to the $B \to S$ transition form factors}

\author{Hua-Yong Han}
\author{Xing-Gang Wu}
\email{wuxg@cqu.edu.cn}
\author{Hai-Bing Fu}
\author{Qiong-Lian Zhang}
\affiliation{Department of Physics, Chongqing University, Chongqing 401331, P.R. China}
\author{Tao Zhong}
\affiliation{Institute of High Energy Physics, Chinese Academy of Sciences, Beijing 100049, P.R. China}

\date{\today}

\begin{abstract}

We investigate the twist-3 light-cone distribution amplitudes (LCDAs) of the scalar mesons $a_0$, $K^{\ast}_0$ and $f_0$ within the QCD sum rules. The QCD sum rules are improved by a consistent treatment of the sizable $s$-quark mass effects within the framework of the background field approach. Adopting the valence quark component $(\bar{q}_1 q_2)$ as the dominant structure of the scalar mesons, our estimation for their masses are close to the measured $a_0(1450)$, $K^{\ast}_0(1430)$ and $f_0(1710)$. From the sum rules, we obtain the first two non-zero moments of the twist-3 LCDAs $\phi^{s,\sigma}_{a_0}$: $\langle \xi_{s,a_0}^{2(4)} \rangle=0.369 \;(0.245)$ and $\langle \xi_{\sigma,a_0}^{2(4)} \rangle=0.203 \;(0.093)$; those of the twist-3 LCDAs $\phi_{K^*_0}^{s,\sigma}$: $\langle \xi_{s,K^{\ast}_0}^{1(2)} \rangle =0.004\;(0.355)$ and $\langle \xi_{\sigma,K^{\ast}_0}^{1(2)} \rangle =0.018\;(0.207)$; and those of the twist-3 LCDAs $\phi_{f_0}^{s,\sigma}$: $\langle \xi_{s,f_0}^{2(4)} \rangle=0.335 \;(0.212)$ and $\langle \xi_{\sigma,f_0}^{2(4)} \rangle=0.196 \; (0.088)$, respectively. As an application of those twist-3 LCDAs, we study the $B \to S$ transition form factors by introducing proper chiral currents into the correlator, which is constructed such that the twist-3 LCDAs give dominant contribution and the twist-2 LCDAs make negligible contribution. Our results of the $B \to S$ transition form factors at the large recoil region $q^2 \simeq 0$ are consistent with those obtained in the literature, which inversely shows the present twist-3 LCDAs are acceptable. \\


\end{abstract}

\maketitle

\section{Introduction}

Even though lots of works have been done in the literature, the properties of the light scalar mesons are still in ambiguity. In order to get an accurate theoretical prediction on the properties of the scalar mesons and on their applications to high energy processes, it is very important to provide a good interpretation of their complicated nonperturbative nature.

Among the scalar mesons' non-perturbative sources, one of the most important thing is their light-cone distribution amplitudes (LCDAs). At the present, some pioneering works for both the twist-2 and twist-3 LCDAs of the scalar meson have been done within the QCD sum rules, c.f. Refs.\cite{cheng2006,lv0612210}. According to our experience on the light pseudoscalar twist-3 LCDAs, e.g. the pion and kaon electromagnetic form factors~\cite{pion,kaon} and the $B\to\pi$, $K$ transition form factors~\cite{bpi,bk}, a well-behaved pseudoscalar twist-3 LCDAs in the end-point region can give the conventional power suppressed contributions to the high energy processes in comparison to those of twist-2 LCDAs. It is interesting to know whether the scalar twist-3 LCDAs also possess such good feature. Moreover, a better understanding of the twist-3 LCDAs is crucial for a reliable estimation. The forthcoming more precise data, e.g. at the large hadronic colliders and the programming super $B$ factories, also requires a more accurate theoretical estimation for the twist-3 contributions.

At present, we will investigate the twist-3 LCDAs of the scalar mesons by incorporating such quark mass effects properly so as to achieve a more accurate theoretical prediction. Within the QCD sum rules, it has been found that the contributions from the quark mass terms (especially those of the $s$-quark) will be comparable to that of the dimension-six operators or even the dimension-four operators, and it can even change the relative importance of the operator expansion series counted by the naive power counting rules, c.f. Refs.\cite{zhong2,zhong,kaon1,kaon2,kaon3,kaon4,kaon5,kaon6} for studying the $SU(3)$-breaking effects of the kaon LCDAs and Refs.\cite{vec1,vec2,vec3} for the cases of the vector twist-3 LCDAs.

For the purpose, we will calculate the Gegenbauer moments of the twist-3 LCDAs for the scalars $a_0$, $K^*_0$ and $f_0$ within the QCD sum rules together with the QCD background field approach. Basic assumption of QCD sum rules is the introducing of nonvanishing vacuum condensates such as the dimension-three quark condensate $\left<\bar{q}q\right>$, the dimension-four gluon condensate $\left<G^2\right>$, and etc.~\cite{shifman}. The QCD background field approach provides a systematic description for those vacuum condensates from the viewpoint of field theory~\cite{BG1,BG2,BG3,BG31,BG4,BG41,BG5,BG6,BG_HT,pro_func}. It assumes that the quark and gluon fields are composed of the background fields and the quantum fluctuations around them. The vacuum expectation values of those background fields describe the nonperturbative effects, while the quantum fluctuations represent the calculable perturbative effects. To take the QCD background field theory as the starting point for the QCD sum rules, it not only shows a distinct physical picture but also greatly simplifies the calculation due to its capability of adopting different gauge conditions for quantum fluctuations and background fields respectively. Because of the influence from background fields, the quark and gluon propagators shall include nonperturbative component inevitably, and the quark mass effect can be introduced in a consistent way.

Moreover, the $B \to S$ transition form factor within the light-cone sum rules (LCSR) provides a good platform for checking the properties of the scalar LCDAs. In the LCSR approach, a two-point correlation function is introduced and expanded near the light cone $x^2=0$, whose matrix elements are parameterized as LCDAs of increasing twists. By using the conventional currents in the correlator, the form factors will always contain the twist-2 and twist-3 terms simultaneously, both of which play important roles for the final LCSRs~\cite{cheng2006}. Because both the twist-2 and twist-3 LCDAs have their own uncertainties, the entanglement of them make the estimation under large uncertainty. Thus, for the sake of a better accuracy, it is helpful to choose proper chiral currents in the correlators such that either the twist-2 or the twist-3 terms make no contribution to the LCSRs. In Ref.\cite{Sun:prd83}, a chiral current has been suggested to make the twist-3 terms give zero contribution. At present, we shall introduce another type of chiral current such that the twist-2 terms make no contributions. Furthermore, we will deal with the semileptonic decays $B\to S l\bar{\nu}_l$ and $B\to S l\bar{l}$, which when in comparing with the forthcoming data shall be helpful for acquiring valuable information on the twist-3 LCDAs of the scalar particles.

The remaining parts of the paper are organized as follows. In Sec.II, we present the calculation technology for deriving the sum rules for the twist-3 LCDA moments of the scalar mesons $a_0$, $K^{\ast}_0$ and $f_0$. Then, we present the formulas for the $B \to S$ transition form factors. Numerical results and discussions for the scalar mesons' twist-3 LCDAs and the $B \to S$ transition form factors are given in Sec.III. The final section is reserved for a summary. In the appendix, we put the some more subtle points in deriving the sum rules for the twist-3 LCDA moments.

\section{Calculation Technology}

\subsection{Twist-3 LCDAs of the Scalar Meson}

\begin{figure}[htb]
\includegraphics[width=0.45\textwidth]{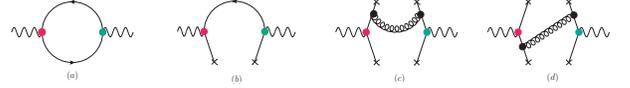}
\caption{Feynman diagrams for calculating the moments of the scalar meson's twist-3 LCDAs $\phi_{S}^{s}$ and $\phi_{S}^{\sigma}$, where the background gluon fields are included in the Fermion propagators implicitly and the background quark fields are depicted as crosses. The left big dot stands for the vertex $\left(i z\cdot \tensor{D}^n \right)$ and $\sigma_{\mu\nu} \left(i z\cdot \tensor{D}^{(n+1)}\right)$ for $\phi_{S}^{s}$ and $\phi_{S}^{\sigma}$ respectively. } \label{tree}
\end{figure}

We adopt the suggestion of the valence quark contents dominant for the scalar mesons $a_0$, $K^{\ast}_0$ and $f_0$, which are $(\bar{d} u)$, $(\bar{u} s)$ and $(\bar{s} s)$ respectively. The twist-3 LCDAs of the scalar meson are defined as~\cite{cheng2006}
\begin{widetext}
\begin{eqnarray}
\langle S(p)|\bar{q}_2(y)q_1(x)|0\rangle &=& m_S \bar{f}_{S} \int_0^1 du \, e^{i(u p\cdot y+ \bar{u}p \cdot x)} \phi_{S}^{s}(u)  \ , \label{eq:lcdasdf}\\
\langle S(p)|\bar{q}_2(y) \sigma_{\mu \nu} q_1(x)|0\rangle &=& -m_{S}(p_{\mu}z_{\nu} -p_{\nu}z_{\mu})  \bar{f}_{S} \int_0^1 du \, e^{i(u p\cdot y+ \bar{u}p \cdot x)} \frac{\phi_{S}^{\sigma}(u)}{6} ,
\label{eq:lcdatdf}
\end{eqnarray}
\end{widetext}
where $z=y-x$, $m_S$ and $p$ are mass and momentum of the scalar meson, $\bar{f}_{S}$ is the decay constant of the scalar meson defined by $\langle S(p)|\bar{q}_{2}q_{1}|0\rangle = m_{S}\bar{f}_{S}$, $u$ is the momentum fraction carried by $q_2$ quark, and $\bar{u}=1-u$. The moments of twist-3 LCDAs are defined as
\begin{eqnarray}
\langle \xi_s^n\rangle &=& \int_0^1 du (2u-1)^n \phi_S^s(u), \label{eq:xidf0} \\
\langle \xi^n_{\sigma}\rangle &=& \int_0^1 du (2u-1)^n \phi_S^{\sigma}(u), \label{eq:xidf1}
\end{eqnarray}
which satisfy
\begin{eqnarray}
&& \langle 0|\bar{q}_1(0)(i z \cdot \buildrel\leftrightarrow\over D)^n q_2(0)|S(p)\rangle = m_S \bar{f}_{S}(p \cdot z)^n  \langle \xi_s^n \rangle , \nonumber \\
&&\langle 0|\bar{q}_1(0)(i z \cdot \buildrel\leftrightarrow\over
D)^{n+1} \sigma_{\mu \nu} q_2(0)|S(p)\rangle = \nonumber\\
&& \quad\quad\quad\quad -i \frac{n+1}{  3}m_S \bar{f}_{S} (p_{\mu}z_{\nu}-p_{\nu}z_{\mu})(p \cdot z)^n \langle \xi_{\sigma}^n \rangle.
\label{eq:moments}
\end{eqnarray}
Here the zeroth moments have been normalized to one, $\langle \xi_s^0\rangle=\langle \xi^0_{\sigma}\rangle=1$.

Fig.(\ref{tree}) shows the Feynman diagrams for calculating the LCDA moments, where the background gluon fields are included in the fermion propagators implicitly and the background quark fields are depicted as crosses.

To study the properties of these LCDAs, one can introduce the following two correlation functions,
\begin{widetext}
\begin{eqnarray}
i \int d^4 x e^{i q \cdot x }\langle 0 |T\{\bar{q}_1(x)(i z \cdot \buildrel\leftrightarrow\over D)^n q_2(x), \bar{q}_2(0) q_1(0)\}|0 \rangle &=& -(z \cdot q)^n I_s^{(n,0)}(q^2),\nonumber \\
i \int d^4 x e^{i q \cdot x }\langle 0 |T\{\bar{q}_1(x)\sigma_{\mu \nu}(i z \cdot \buildrel\leftrightarrow\over D)^{n+1} q_2(x), \bar{q}_2(0) q_1(0)\}|0\rangle &=& i (q_{\mu}z_{\nu}-q_{\nu}z_{\mu})(z \cdot q)^n I_{\sigma}^{(n,0)} (q^2).
\end{eqnarray}
\end{widetext}

Following the standard QCD LCSR within the background field theory \cite{BG1,BG2,BG3,BG4,BG_HT,pro_func}, we can derive the sum rules for the moments of $\phi_{S}^{s}$ and $\phi_{S}^{\sigma}$. For convenience, we present the detailed processes in Appendix A.

During the calculation, as has been argued by Ref.\cite{zhong}, we should deal with the sizable $s$-quark mass effects in a more consistent way, which might cause sizable effects comparable to those of the dimension-six condensates. For the purpose, we adopt the following propagators to do our calculation~\cite{zhong}
\begin{widetext}
\begin{eqnarray}
S_{F}(x,0) &=& i \int \frac{d^4 q}{(2\pi)^4} e^{-iq \cdot x}
 \bigg \{ -\frac{m + \not\! q}{m^2 - q^2} + \frac{\gamma^\nu ( \not\! q - m )
\gamma^\mu}{(m^2 - q^2)^2} b_{0\nu\mu} - i \bigg [ 2 \frac{\gamma^\nu (\not\! q - m) q^\rho}
{(m^2 - q^2)^3} + \frac{g^{\nu\rho}}{(m^2 - q^2)^2} \bigg]
\gamma^\mu b_{1\nu\mu|\rho} +\cdots \bigg\} \label{eq:propagatorofquark}
\end{eqnarray}
\end{widetext}
for the quark propagator and
\begin{equation}
S^{ab}_{\mu\nu}(x,0)=i \int \frac{d^4 q}{(2\pi)^4} e^{-iq \cdot x}
\left\{-\frac{g_{\mu\nu}}{q^2} \delta^{ab} +\cdots \right\}  \label{eq:propagatorofgluon}
\end{equation}
for the gluon propagator, where $b_{0\nu\mu}=\frac{i}{2}G_{\nu\mu}(0)$ and $b_{1\nu\mu|\rho}
= \frac{i}{3} \left[ G_{\nu\mu;\rho}(0)+ G_{\rho\mu;\nu}(0)\right]$. Here $G_{\nu\mu}(x) = g_s T^a G^a_{\nu\mu}(x)$, the gauge invariant function $G_{\nu\mu;\rho}(0)=g_s T^a \widetilde{D}^{ab}_{\rho} G^{b}_{\nu\mu}(x)|_{x=0}$, and the symbol $\cdots$ stands for the irrelevant terms that lead to higher-order operators over than dimension-six.

The sum rule for the even moments of $\phi_{S}^{s}$ up to dimension-six condensates is
\begin{widetext}
\begin{eqnarray}
&&-m_S^2 \bar{f}_S^2 e^{-m_S^2/M^2}\langle \xi^{2n}_s\rangle \nonumber \\
&=& + \frac{3}{ 4 \pi^2}\int^{1}_0(2x-1)^{2n}\left [-(2n+3)x(1-x)+\frac{m_1 m_2-m_{12}^2}
{ M^2}\right] M^4e^{-\frac{m_{12}^2 }{ M^{2}x(1-x)}}dx \nonumber  \\
&&-\frac{3}{ 4 \pi^2}\int^{1}_0 (2x-1)^{2n}\left[-(2n+3)x(1-x)\left(1+\frac{S_s}{ M^2}\right)
+\frac{2(n+1)m_{12}^2 + m_1 m_2}{ M^2}\right] M^4e^{-{S_s/M^2}}dx \nonumber \\
&&+\langle \alpha_s G^2\rangle\int^{1}_0(2x-1)^{2n}\frac{1}{ 8\pi}\left[
-(2n+1)+\frac{2m_1m_2-m_{12}^2} {M^2 x(1-x)}\right] e^{-\frac{m_{12}^2 }{ M^{2}x(1-x)}}dx \nonumber  \\
&&+\langle g_s^3fG^3\rangle \frac{n(2n-1)}{48\pi^2}\int^{1}_0(2x-1)^{2n-2} \frac{1}{M^2}e^{-\frac{m_{12}^2 }{ M^{2}x(1-x)}}dx  \nonumber  \\
&&+\bigg \{ -\langle \bar{q}_1 q_1\rangle \left[ \frac{2m_2+(2n+1)m_1}{2}+
\frac{2n(4n+1)m_1^3+12nm_1^2m_2+3m_1m_2^2}{6M^2} +\frac{m_1^2m_2^2(2nm_1+m_2)} {2M^4}+\frac{m_1^3m_2^4}{6M^6}\right]e^{-m_2^2/M^2}  \nonumber  \\
&&+\langle g_s\bar{q}_1\sigma TGq_1\rangle \bigg [ \frac{9(2n-1)m_2
+n(16n-5)m_1}{18M^2}+\frac{((8n-3)m_1+3m_2)m_2^2}{12M^4}+\frac{m_1m_2^4}
{9M^6}\bigg ]e^{-m_2^2/M^2}\nonumber  \\
&&+g_s^2 \langle \bar{q}_1q_1\rangle^2 \left[\frac{-8n^2-14n+12}
{81M^2}-\frac{m_2^2}{27M^4}\right]e^{-m_2^2/M^2} + g_s^2 \langle \bar{q}_1q_1\rangle \langle \bar{q}_2q_2 \rangle \frac{4}{9}\left[\frac{2e^{-m_2^2/M^2}}{M^2}+\frac{1}{m_2^2}
(e^{-m_2^2/M^2}-1)\right] \nonumber \\
&&+g_s^2 \langle \bar{q}_1q_1\rangle \langle \bar{q}_2q_2\rangle
 \frac{2}{9}\frac{1}{m_2^2-m_1^2}\left[(e^{-m_1^2/M^2}-e^{-m_2^2/M^2}) +2m_1 m_2 \left(\frac{e^{-m_1^2/M^2}-1}{m_1^2}-\frac{e^{-m_2^2/M^2}-1}{m_2^2}\right)\right]\nonumber \\
&& +\left[q_1 \leftrightarrow q_2, m_1 \leftrightarrow m_2 \right] \bigg \} ,
\label{eq:resultslcdase}
\end{eqnarray}
\end{widetext}
where $m_{12}^2=m_{1}^{2} x + m_{2}^{2} (1-x)$, $m_{1}$ and $m_{2}$ denote the masses of the $q_{1}$ and $\bar{q}_{2}$ quarks, $M$ is the Borel parameter.

The sum rule for the even moments of $\phi_{S}^{\sigma}$ up to dimension-six condensates is
\begin{widetext}
\begin{eqnarray}
&&-\frac{1}{3}m_S^2 \bar{f}_S^2 e^{-m_S^2/M^2}\langle \xi^{2n}_{\sigma}\rangle \nonumber \\
&=& \frac{3}{4\pi^2}\int^1_0dx(2x-1)^{2n}M^4x(1-x)e^{-\frac{m_{12}^2}{M^{2}x(1-x)}} -\frac{3}{4 \pi^2}\int^1_0dx (2x-1)^{2n}\left\{x(1-x)\left(1+\frac{S_{\sigma}}
{ M^2}\right) +\frac{m_{12}^2}{M^2}\right\} M^4 e^{-{S_{\sigma}/M^2}} \nonumber \\
&&-\langle \alpha_s G^2\rangle \int^1_0dx\frac{(2x-1)^{2n}}{24\pi}
\left\{ 1-\frac{2m_1m_2}{M^2x(1-x)} \right\}e^{-\frac{m_{12}^2}
{M^{2}x(1-x)}}+\langle g_s^3fG^3\rangle \frac{m_1m_2}{24\pi^2}\int^1_0dx
\frac{(2x-1)^{2n}}{2M^4x(1-x)}e^{-\frac{m_{12}^2}{M^2x(1-x)}}\nonumber \\
&&+\bigg \{\langle \bar{q}_1q_1\rangle \frac{-1}{6} \left[ 3m_1+\frac{(4n+1)m_1^3}
{M^2}+\frac{m_1^3m_2^2}{M^4}\right] e^{-m_2^2/M^2}+\langle g_s \bar{q}_1 \sigma TGq_1\rangle \bigg [
\frac{(16n+1)m_1+6m_2}{36M^2}+\frac{m_1m_2^2}{9M^4}\bigg ]e^{-m_2^2/M^2}\nonumber \\
&&+\frac{g_s^2\langle \bar{q}_1q_1\rangle^2}{81}\left[ \frac{-4n+5}{M^2}+\frac{2m_2^2}{M^4}\right]
 e^{-m_2^2/M^2} +\left[q_1 \leftrightarrow q_2, m_1 \leftrightarrow m_2\right] \bigg \}.
\label{eq:resultslcdate}
\end{eqnarray}
\end{widetext}

The sum rule for the odd moments of $\phi_{S}^{s}$ up to dimension-six condensates is
\begin{widetext}
\begin{eqnarray}
&&-m_S^2 \bar{f}_S^2 e^{-m_S^2/M^2}\langle \xi^{2n+1}_s\rangle \nonumber \\
&=& \frac{3}{ 4 \pi^2}\int^{1}_0(2x-1)^{2n+1}\left[-2(n+2)x(1-x)+\frac{m_1m_2-m_{12}^2}
{ M^2}\right] M^4e^{-\frac{m_{12}^2 }{ M^{2}x(1-x)}}dx \nonumber \\
&&-\frac{3}{ 4 \pi^2}\int^{1}_0 (2x-1)^{2n+1}\left[-2(n+2)x(1-x)\left(1+\frac{S_s}{ M^2}\right) +\frac{(2n+3)m_{12}^2 + m_1 m_2} { M^2}\right] M^4 e^{-{S_s/M^2}}dx \nonumber \\
&&+\langle \alpha_s G^2\rangle\int^{1}_0(2x-1)^{2n+1}\frac{1}{ 8\pi}\left[
-2(n+1)+\frac{2m_1m_2-m_{12}^2}{M^2 x(1-x)}\right] e^{-\frac{m_{12}^2}{ M^{2}x(1-x)}}dx \nonumber  \\
&&+\langle g_s^3fG^3 \rangle \frac{2n(2n+1)}{96\pi^2}\int^{1}_0(2x-1)^{2n-1}
\frac{1}{M^2}e^{-\frac{m_{12}^2}{ M^{2}x(1-x)}}dx  \nonumber  \\
&&+\bigg \{-\langle \bar{q}_1q_1\rangle \bigg [ \frac{2m_2+2(n+1)m_1}{2}+ \frac{(2n+1)(4n+3)m_1^3 +6(2n+1)m_1^2m_2+3m_1m_2^2}{6M^2}+\frac{m_1^2m_2^2((2n+1)m_1+m_2)} {2M^4} \nonumber \\
&&+\frac{m_1^3m_2^4}{6M^6}\bigg ]e^{-m_2^2/M^2} +\langle g_s\bar{q}_1\sigma T Gq_1\rangle \left[ \frac{36nm_2 +(2n+1)(16n+3)m_1}{36M^2}+\frac{((8n+1)m_1+3m_2)m_2^2}{12M^4}+\frac{m_1m_2^4}
{9M^6}\right] e^{-m_2^2/M^2}\nonumber  \\
&&+g_s^2 \langle \bar{q}_1q_1\rangle^2\bigg [\frac{-2(2n-1)^2-14n+5}
{81M^2}-\frac{m_2^2}{27M^4}\bigg ]e^{-m_2^2/M^2}+4\pi \alpha_s \langle \bar{q}_1q_1\rangle \langle \bar{q}_2q_2 \rangle \frac{4}{9}\left[\frac{2e^{-m_2^2/M^2}}{M^2}+\frac{1}{m_2^2}
(e^{-m_2^2/M^2}-1)\right] \nonumber \\
&&+4\pi \alpha_s \langle \bar{q}_1q_1\rangle \langle \bar{q}_2q_2\rangle
 \frac{2}{9}\frac{1}{m_2^2-m_1^2}\left[(e^{-m_1^2/M^2}-e^{-m_2^2/M^2}) +2m_1 m_2 \left(\frac{e^{-m_1^2/M^2}-1}{m_1^2}-\frac{e^{-m_2^2/M^2}-1}{m_2^2}\right)\right] \nonumber \\
&& -\left[q_1 \leftrightarrow q_2, m_1 \leftrightarrow m_2 \right] \bigg \}.
\label{eq:resultslcdaso}
\end{eqnarray}
\end{widetext}

The sum rule for the odd moments of $\phi_{S}^{\sigma}$ up to dimension-six condensates is
\begin{widetext}
\begin{eqnarray}
&&-\frac{1}{3}m_S^2 \bar{f}_S^2 e^{-m_S^2/M^2}\langle \xi^{2n+1}_{\sigma}\rangle \nonumber \\
&=& \frac{3}{4\pi^2}\int^1_0dx(2x-1)^{2n+1}M^4x(1-x)e^{-\frac{m^2_{12}}{M^{2}x(1-x)}} -\frac{3}{4 \pi^2}\int^1_0dx (2x-1)^{2n+1}\left[x(1-x)\left(1+\frac{S_{\sigma}} { M^2}\right) +\frac{m^2_{12}}{M^2}\right] M^4 e^{-{S_{\sigma}/M^2}} \nonumber  \\
&&-\langle \alpha_s G^2\rangle \int^1_0dx\frac{(2x-1)^{2n+1}}{24\pi}
\left[ 1-\frac{2m_1m_2}{M^2x(1-x)} \right] e^{-\frac{m^2_{12}}
{M^{2}x(1-x)}} +\langle g_s^3fG^3\rangle \frac{m_1m_2}{24\pi^2}\int^1_0dx
\frac{(2x-1)^{2n+1}}{2M^4x(1-x)}e^{-\frac{m^2_{12}}{M^2x(1-x)}}\nonumber
\\
&&+\bigg \{\langle \bar{q}_1q_1\rangle \frac{-1}{6}\bigg [ 3m_1+\frac{(4n+3)m_1^3}
{M^2}+\frac{m_1^3m_2^2}{M^4}\bigg ] e^{-m_2^2/M^2}
+\langle g_s \bar{q}_1 \sigma TGq_1\rangle \bigg [
\frac{(16n+9)m_1+6m_2}{36M^2}+\frac{m_1m_2^2}{9M^4}\bigg ]e^{-m_2^2/M^2}\nonumber \\
&&+\frac{g_s^2\langle \bar{q}_1q_1\rangle^2}{81}\bigg [ \frac{-4n+1}{M^2}+\frac{2m_2^2}{M^4}\bigg ]
 e^{-m_2^2/M^2}
-\left[q_1 \leftrightarrow q_2, m_1 \leftrightarrow m_2\right] \bigg \}.
\label{eq:resultslcdato}
\end{eqnarray}
\end{widetext}

%%%%%%%%%%%%%%%%%%%%%%%%%%%%%%%%%%%%%%%%%%%%%%%%%%%%%%%%%%%%%%%%%%%%%%%%%
\subsection{$B \to S$ Transition Form Factors within the Light-Cone Sum Rules}
\label{ffd}
%%%%%%%%%%%%%%%%%%%%%%%%%%%%%%%%%%%%%%%%%%%%%%%%%%%%%%%%%%%%%%%%%%%%%%%%%

The $B \to S$ transition form factors $f_{\pm}(q^2)$ and $f_{T}(q^2)$ are defined through the hadronic matrix elements $\langle S(p) | \bar {q_2}\gamma_{\mu}\gamma_5b | B_{(s)}(p+q)\rangle$ and $\langle S(p) | \bar{q_2} \sigma_{\mu\nu}\gamma_5q^\nu b| B_{(s)}(p+q)\rangle$, which are
\begin{widetext}
\begin{eqnarray}
 \langle S(p)|\bar q_2\gamma_{\mu}\gamma_5b| B_{(s)}(p+q)\rangle  &=&
-2ip_\mu f_{+}(q^2)-i[f_{+}(q^2)+f_{-}(q^2)]q_\mu ,  \label{eq:formfdf1} \\
\langle S(p)|\bar q_2\sigma_{\mu\nu} \gamma_5q^\nu b|  B_{(s)}(p+q)\rangle
&=& [2p_\mu q^2-2q_\mu (p\cdot q)]\frac{-f_T(q^2)}{m_{B_{(s)}}+m_S}. \label{eq:formfdf2}
\end{eqnarray}
\end{widetext}
These form factors are key factors for studying the semileptonic decays $B_{(s)}\to Sl\bar{\nu}_l$ and $B_{(s)}\to Sl\bar{l}$.

By introducing proper chiral correlators, we obtain the light-cone sum rules for those form factors that depend only on the twist-3 LCDAs of the scalar mesons, and then our twist-3 LCDAs derived in the last subsection apply. More explicitly, we suggest to calculate the following correlators :
\begin{widetext}
\begin{eqnarray}
\Pi_{\mu}(p,q)&=& i \int d^4 xe^{iqx} \langle S(p)|T\{
\bar{q_2}(x)\gamma_{\mu}(1-\gamma_5)b(x),\bar{b}(0)i(1+\gamma_5)q_1(0)\}
|0 \rangle, \label{eq:correlationfunctionfpm}\\
\widetilde{\Pi}_{\mu}(p,q)&=& i \int d^4 xe^{iqx} \langle S(p)|T\{
\bar{q_2}(x)\sigma_{\mu\nu}(1+\gamma_5)q^{\nu}b(x),\bar{b}(0)i(1+\gamma_5)q_1(0)\}
|0 \rangle. \label{eq:correlationfunctionft}
\end{eqnarray}
\end{widetext}
where $q_1,q_2$ denotes the light quark field. Following the standard procedures to deal with the correlators which are similar to that of the $B\to$ pseudoscalar transition form factors, c.f. Refs.\cite{Huang:prd63,Zhou:0402,BKwu1,BKwu2,Khodjamirian:hep15}, we can obtain the light-cone sum rules for the form factors $f_{\pm,T}(q^2)$:
\begin{widetext}
\begin{eqnarray}
f_+(q^2) &=& \frac{m_b + m_{q_1}}{m_{B_{q_1}}^2 f_{B_{q_1}}}\exp \left(\frac{m_{B_{q_1}}^2}{M^2}\right) \bigg \{ \int^1_{u_0}\frac{du}{u}\exp \left[{-\frac{m_b^2+u \bar{u}p^2-\bar{u}q^2}{uM^2}}\right] \times \bigg[m_S \bar{f}_{S} ( u \phi_S^s(u)+\frac{1}{3}\phi_S^{\sigma}(u)) \nonumber \\
&& +\frac{m_S \bar{f}_{S}}{6uM^2}  \phi_S^{\sigma}(u)(m_b^2-u^2p^2+q^2)\bigg]
+\frac{m_S \bar{f}_{S}}{6} \phi_S^{\sigma}(u_0)\exp \left(-\frac{s_0}{M^2}\right) \frac{m_b^2-u_0^2p^2+q^2}{m_b^2+u_0^2p^2-q^2} \bigg \},  \label{results formfactors plus} \\
f_+(q^2)+f_-(q^2) &=& 2\frac{m_b+m_{q_1}}{m_{B_{q_1}}^2 f_{B_{q_1}}} \exp \left(\frac{m_{B_{q_1}}^2}{M^2}\right)
\bigg \{ \int_{u_0}^1 \frac{du}{u} \exp \left[-\frac{m_b^2+u \bar{u}p^2-\bar{u}q^2}{uM^2}\right] \times \bigg[m_S \bar{f}_{S}(\phi_S^s(u)+\frac{1}{6u}\phi_S^{\sigma}(u)) \nonumber \\
&&-\frac{m_S \bar{f}_{S}}{6u^2M^2}\phi_S^{\sigma}(u)(m_b^2+u^2p^2-q^2)\bigg]
-\frac{m_S \bar{f}_{S}}{6u_0}\phi_S^{\sigma}(u_0) \exp \left(-\frac{s_0}{M^2}\right) \bigg \}, \label{results formfactors minus} \\
f_T(q^2) &=& \frac{(m_b+m_{q_1})(m_{B_{q_1}}+m_S)}{m_{B_{q_1}}^2f_{B_{q_1}}}\exp \left(\frac{m_{B_{q_1}}^2}{M^2}\right) \bigg\{ \int_{u_0}^1 \frac{du}{u}\exp \left[-\frac{m_b^2+u\bar{u}p^2-\bar{u}q^2}{uM^2}\right] \times \frac{m_b m_S \bar{f}_{S}}{3uM^2}\phi_S^{\sigma}(u) \nonumber \\
&&+\frac{m_b m_S \bar{f}_{S}}{3} \phi_S^{\sigma}(u_0)\exp \left(-\frac{s_0}{M^2}\right)
\frac{1}{m_b^2+u_0^2p^2-q^2} \bigg \}. \label{results formfactors T}
\end{eqnarray}
\end{widetext}
with
\begin{displaymath}
u_0=\frac{\sqrt{(s_0-q^2-p^2)^2+4p^2(m_b^2-q^2)}-(s_0-q^2-p^2)}{2p^2}.
\end{displaymath}

%%%%%%%%%%%%%%%%%%%%%%%%%%%%%%%%%%%%%%%%%%%%%%%%%%%%%%%%%%%%%%%%
\section{Numerical Results and Discussions}
\label{nrlcdas}
%%%%%%%%%%%%%%%%%%%%%%%%%%%%%%%%%%%%%%%%%%%%%%%%%%%%%%%%%%%%%%%%

We adopt the following input parameters to do our numerical analysis \cite{pdg,5070}:
\begin{eqnarray}
&& m_u=(2.5\pm 0.8) \ {\rm{MeV}}, m_d=(5.0\pm 0.8) \ {\rm{MeV}}, \\
&& m_s=(101\pm 12) \ {\rm{MeV}}, \alpha_{s}(m_c=1.27 \ {\rm{GeV}})=0.39
\end{eqnarray}
and
\begin{eqnarray}
\langle \alpha_s G^2 \rangle &=& (7.5 \pm 2)\times10^{-2} \ {\rm{GeV}}^4, \nonumber\\
\langle g_s^3fG^3 \rangle &=& (8.2 \pm 1) \times \langle \alpha_s  G^2 \rangle, \nonumber\\
\langle \bar{u} u\rangle &\cong& \langle \bar{d} d\rangle \cong -(0.254 \pm 0.015)^3 {\rm{GeV}}^3 , \nonumber\\
\langle \bar{s}s\rangle &=& (0.74\pm 0.03)\times \langle \bar{u}u\rangle, \nonumber\\
\langle g_s\bar{u}\sigma TG u\rangle &\cong& \langle g_s\bar{d}\sigma T G d\rangle=m_0^2 \langle \bar{u}u\rangle, \nonumber\\
m_0^2 &=& (0.80 \pm 0.02) \ {\rm{GeV}}^2, \nonumber \\
g_s^2\langle \bar{u}u\rangle^2 &=& g_s^2 \langle \bar{d}d\rangle^2 = 2.693\times 10^{-3} \ {\rm{GeV}}^6,\nonumber\\
g_s^2\langle \bar{s}s\rangle^2 &=& 0.74\times g_s^2 \langle \bar{u}u\rangle^2 .
\end{eqnarray}
These condensates are given at the energy scale $\mu=2$ GeV, which when necessary, will be run to the required scales by using the evolution equations \cite{yang,Hwang}.

\subsection{Masses and decay constants for the scalar mesons}

Setting $n=0$ in the sum rules (\ref{eq:resultslcdase}, \ref{eq:resultslcdate}) for the moments of $\phi_{S}^{s}$ and $\phi_{S}^{\sigma}$, we can further derive the sum rules for the masses of the scalar mesons. The mass sum rules are derived by doing the logarithm of these equations and applying the differential operation $M^4 \partial/ \partial M^2$ over them. It is found that the results from both sum rules (\ref{eq:resultslcdase}, \ref{eq:resultslcdate}) are consistent with each other, so we adopt the results from sum rule (\ref{eq:resultslcdase}) for a detailed discussion.

\begin{table}[tb]
\caption{Masses for the scalar mesons. The threshold parameters are taken as $S_{s,\sigma}^{a_0}=(6.0\pm0.5) \rm{GeV}^2$, $S_{s,\sigma}^{K^{\ast}_0} =(5.4\pm0.3) \rm{GeV}^2$ and $S_{s,\sigma}^{f_0}=(6.5\pm0.3) \rm{GeV}^2$. }
\begin{center}
\begin{tabular}{|c|c|c|}
\hline\hline
~~~mesons~~~  & ~~~$M^2(\rm{GeV}^2)$~~~   &~~~$m(\rm{GeV})$~~~ \\
\hline
    $a_0$     & $1.037 \sim 1.666$  & $1.312 \sim 1.571$ \\
\hline
$K_0^{\ast}$  & $0.991 \sim 1.501$  & $1.328 \sim 1.514$ \\
\hline
   $f_0$      & $1.284 \sim 1.803$  & $1.563 \sim 1.706$ \\
\hline
\end{tabular}
\end{center}\label{tabmass}
\end{table}

\begin{figure}[tb]
\includegraphics[width=0.4\textwidth]{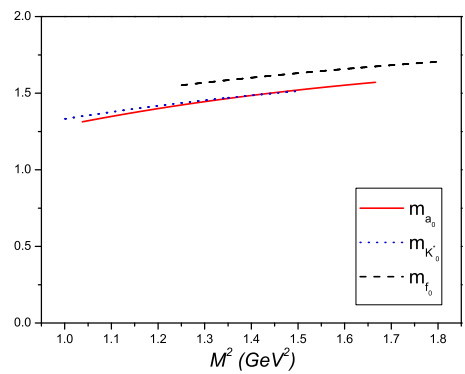}
\caption{Masses of scalar mesons $a_0$, $K^*_0$ and $f_0$ versus $M^2$ within their Borel windows. }\label{figmass}
\end{figure}

Usually, the threshold parameter $S_s$ (or $S_{\sigma}$) is taken to be around the squared mass of the scalar's first excited state. We adopt $S_{s,\sigma}^{a_0}=(6.0\pm0.5) \rm{GeV}^2$ for $a_0$~\cite{cheng2006}, $S_{s,\sigma}^{K^{\ast}_0} =(5.4\pm0.3) \rm{GeV}^2$ and $S_{s,\sigma}^{f_0}=(6.5\pm0.3) \rm{GeV}^2$ for $K^{\ast}_0$ and $f_0$~\cite{lv0612210}. According to the SVZ sum rule, the Borel window for the parameter $M$ is determined by the requirement that the dimension-six condensate contribution (SIX) does not exceed $10\%$ and the continuum contribution (CON) is not too large, i.e. less than $30\%$ of the total dispersive integration. In Table \ref{tabmass} and Fig.(\ref{figmass}), the Borel window is obtained by setting ${\rm SIX}<1\%$ and ${\rm CON}<30\%$. The masses for the scalar mesons are presented in Table \ref{tabmass} and their values versus $M^2$ are presented in Fig.(\ref{figmass}), whose central values are
\begin{displaymath}
m_{a_0}=1442 {\rm MeV}, m_{K^{\ast}_0}=1421 {\rm MeV}, m_{f_0}=1634 {\rm MeV},
\end{displaymath}
These values are close to the physical states $a_0(1450)$, $K^{\ast}_0(1430)$ and $f_0(1710)$ \cite{pdg}. This shows that the valence quark constituent $\bar{d} u$, $\bar{u} s$ and $\bar{s} s$ are viable choices for studying the properties of $a_0$, $K^{\ast}_0$ and $f_0$.

\begin{table}[tb]
\caption{Decay constants for the scalar mesons at the energy scale $\mu=1$ GeV. }
\begin{center}
\begin{tabular}{|c|c|c|}
\hline\hline
~~~mesons~~~   & ~~~$M^2(\rm{GeV}^2)$~~~   & ~~~$\bar{f}(\rm{GeV})$~~~ \\
\hline
$a_0$    &$1.6 \sim 2.5$  &$0.374 \sim 0.377 $  \\
\hline
$K_0^{\ast}$ &$1.5 \sim 2.2$   &$0.357 \sim 0.359$    \\
\hline
$f_0$     &$1.8 \sim 2.2$    &$0.374 \sim 0.378$    \\
\hline
\end{tabular}
\end{center}\label{tabdecayconstant}
\end{table}

\begin{figure}[tb]
\includegraphics[width=0.4\textwidth]{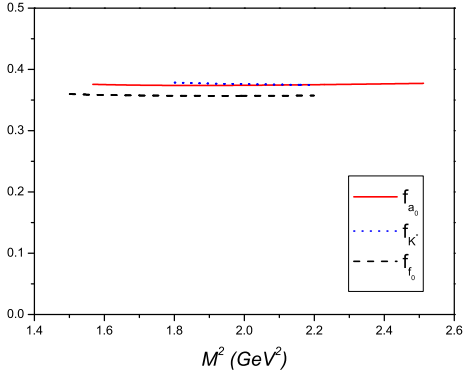}
\caption{Decay constants of scalar mesons $a_0$, $K^*_0$ and $f_0$ versus $M^2$ within their Borel windows. }\label{figdecay}
\end{figure}

Taking the scalar masses as inputs, we can further calculate the decay constants of the scalar mesons. For the purpose, the Borel window for each meson is redetermined following the same criteria as above, whose values are collected in Table \ref{tabdecayconstant}. The decay constants at the energy scale $1$ GeV for the scalar mesons are presented in Table \ref{tabdecayconstant} and their values versus $M^2$ are presented in Fig.(\ref{figdecay}), whose central values are
\begin{displaymath}
\bar{f}_{a_0}=375 {\rm MeV}, \bar{f}_{K^{\ast}_0}=358 {\rm MeV}, \bar{f}_{f_0}=376 {\rm MeV}.
\end{displaymath}

In the following subsections, all the values of decay constants and moments are given at the scale $1$ GeV unless explicitly pointed out.

\subsection{Moments for the scalar mesons}

\begin{table}[htb]
\caption{Moments from the scalar density sum rules.}
\begin{center}
\begin{tabular}{|c|c|c|}
\hline\hline
~~~mesons~~~  & ~~~$\langle \xi_s \rangle$~~~  & ~~~$M^2(\rm{GeV}^2)$~~~    \\
\hline
   &$\langle \xi_{s,a_0}^2\rangle=(0.367\sim0.371)$ &$2.0\sim2.5$ \\
\cline{2-3}
\raisebox{2.0ex}[0pt]{$a_0$}
   &$\langle \xi_{s,a_0}^4\rangle=(0.24\sim0.25)$   &$2.0\sim2.6$ \\
\hline
   &$\langle \xi_{s,k_0^{\ast}}^1\rangle=(0.33\sim0.44)\times10^{-2}$ &$2.0\sim3.5$ \\
\cline{2-3}
\raisebox{2.0ex}[0pt]{$k_0^{\ast}$}
   &$\langle \xi_{s,k_0^{\ast}}^2\rangle=(0.352\sim0.358)$   &$1.8\sim2.3$ \\
\hline
   &$\langle \xi_{s,f_0}^2\rangle=(0.331\sim0.339)$ &$1.8\sim2.2$  \\
\cline{2-3}
\raisebox{2.0ex}[0pt]{$f_0$}
   &$\langle \xi_{s,f_0}^4\rangle=(0.204\sim0.220)$   &$1.8\sim2.2$  \\
\hline
\end{tabular}
\end{center}\label{tabmomentss}
\end{table}

\begin{table}[htb]
\caption{Moments from the tensor sum rules.}
\begin{center}
\begin{tabular}{|c|c|c|}
\hline\hline
~ mesons~   & ~~~$\langle \xi_\sigma \rangle$~~~  & ~~~$M^2(\rm{GeV}^2)$~~~    \\
\hline
   &$\langle \xi_{\sigma,a_0}^2\rangle=(0.203\sim0.204)$ &$2.0\sim2.5$  \\
\cline{2-3}
\raisebox{2.0ex}[0pt]{$a_0$}
   &$\langle \xi_{\sigma,a_0}^4\rangle=(0.092\sim0.094)$   &$2.0\sim2.5$  \\
\hline
   &$\langle \xi_{\sigma,k_0^{\ast}}^1\rangle=(0.99\sim2.54)\times10^{-2}$ &$1.8\sim7.8$ \\
\cline{2-3}
\raisebox{2.0ex}[0pt]{$k_0^{\ast}$}
   &$\langle \xi_{\sigma,k_0^{\ast}}^2\rangle=(0.206\sim0.208)$  &$1.8\sim2.3$  \\
\hline
   &$\langle \xi_{\sigma,f_0}^2\rangle=(0.192\sim0.199)$ &$1.5\sim2.2$  \\
\cline{2-3}
\raisebox{2.0ex}[0pt]{$f_0$}
   &$\langle \xi_{\sigma,f_0}^4\rangle=(0.085\sim0.091)$   &$1.8\sim2.2$  \\
\hline
\end{tabular}
\end{center}\label{tabmomentst}
\end{table}

\begin{figure}
\includegraphics[width=0.22\textwidth]{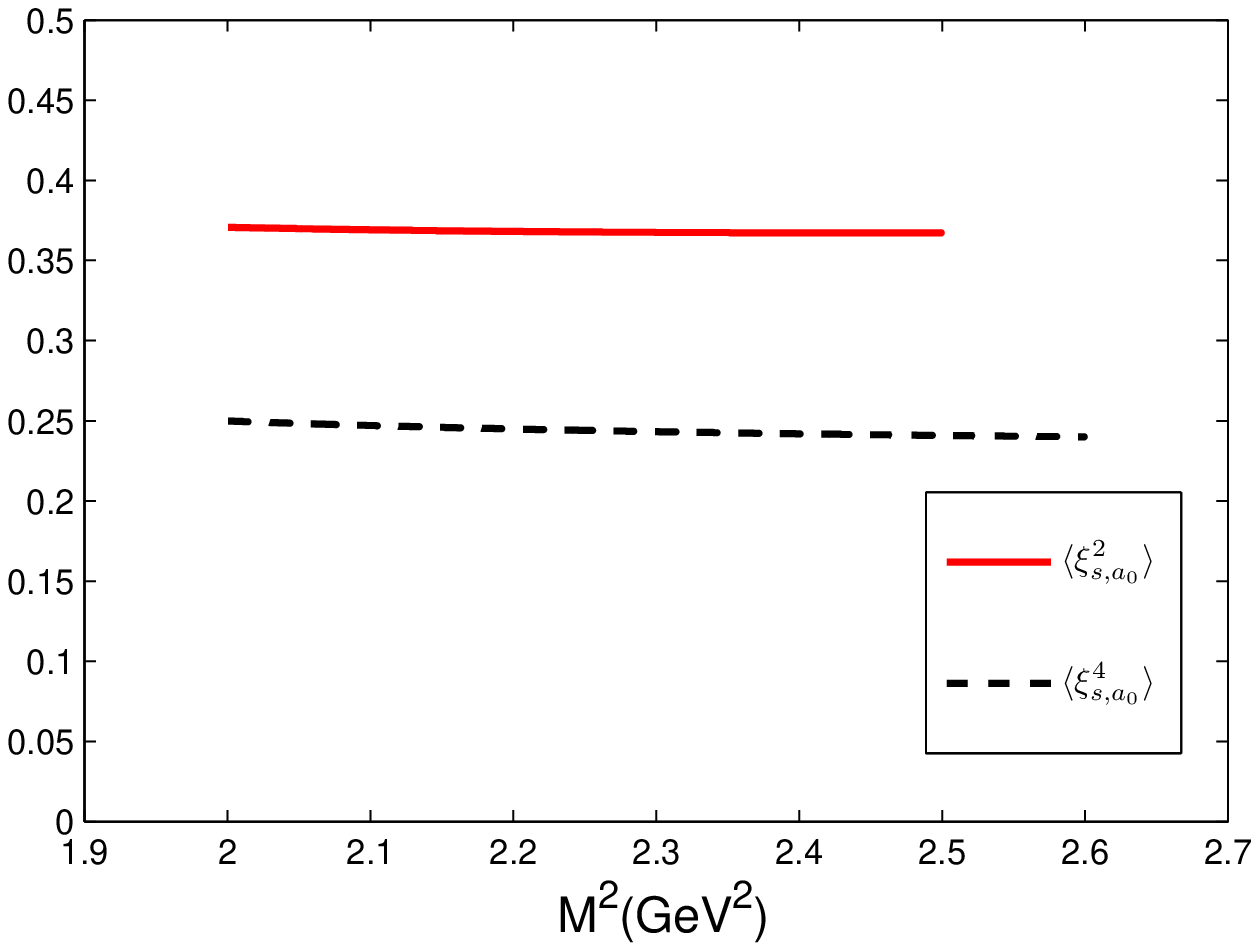}
\includegraphics[width=0.22\textwidth]{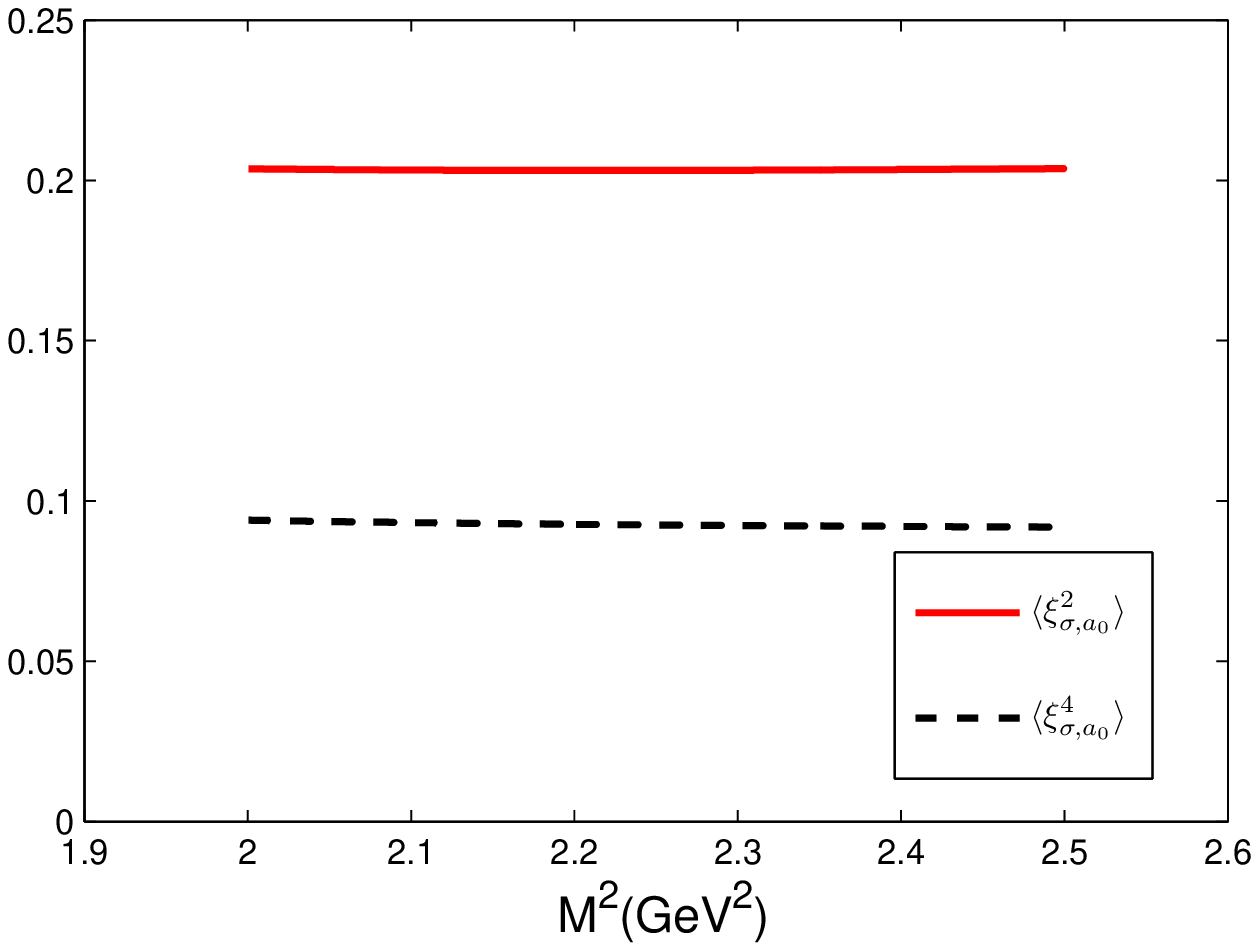}
\caption{Moments of scalar meson $a_0$ versus $M^2$ within their Borel windows.
}\label{figa0xi}
\end{figure}

\begin{figure}
\includegraphics[width=0.22\textwidth]{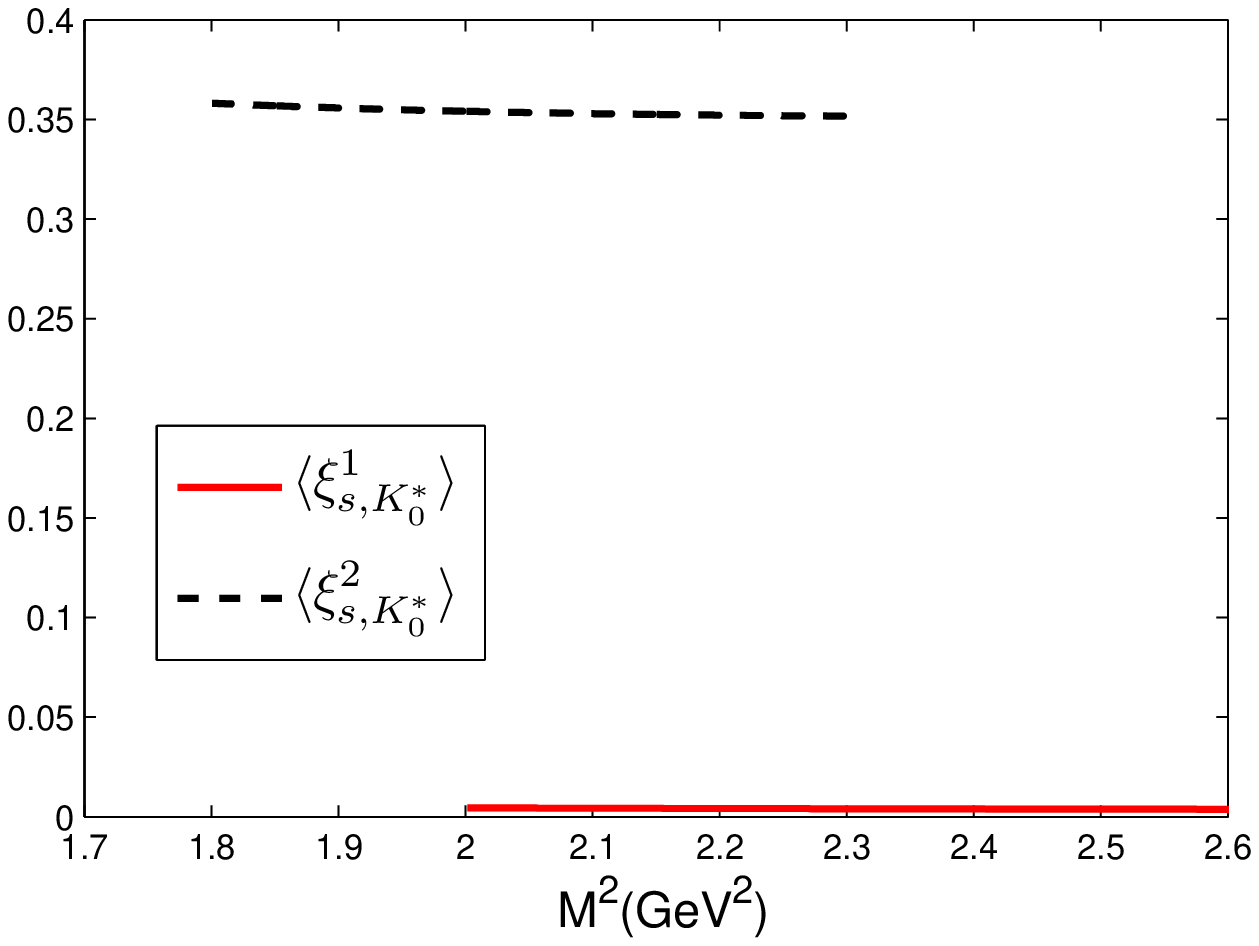}
\includegraphics[width=0.22\textwidth]{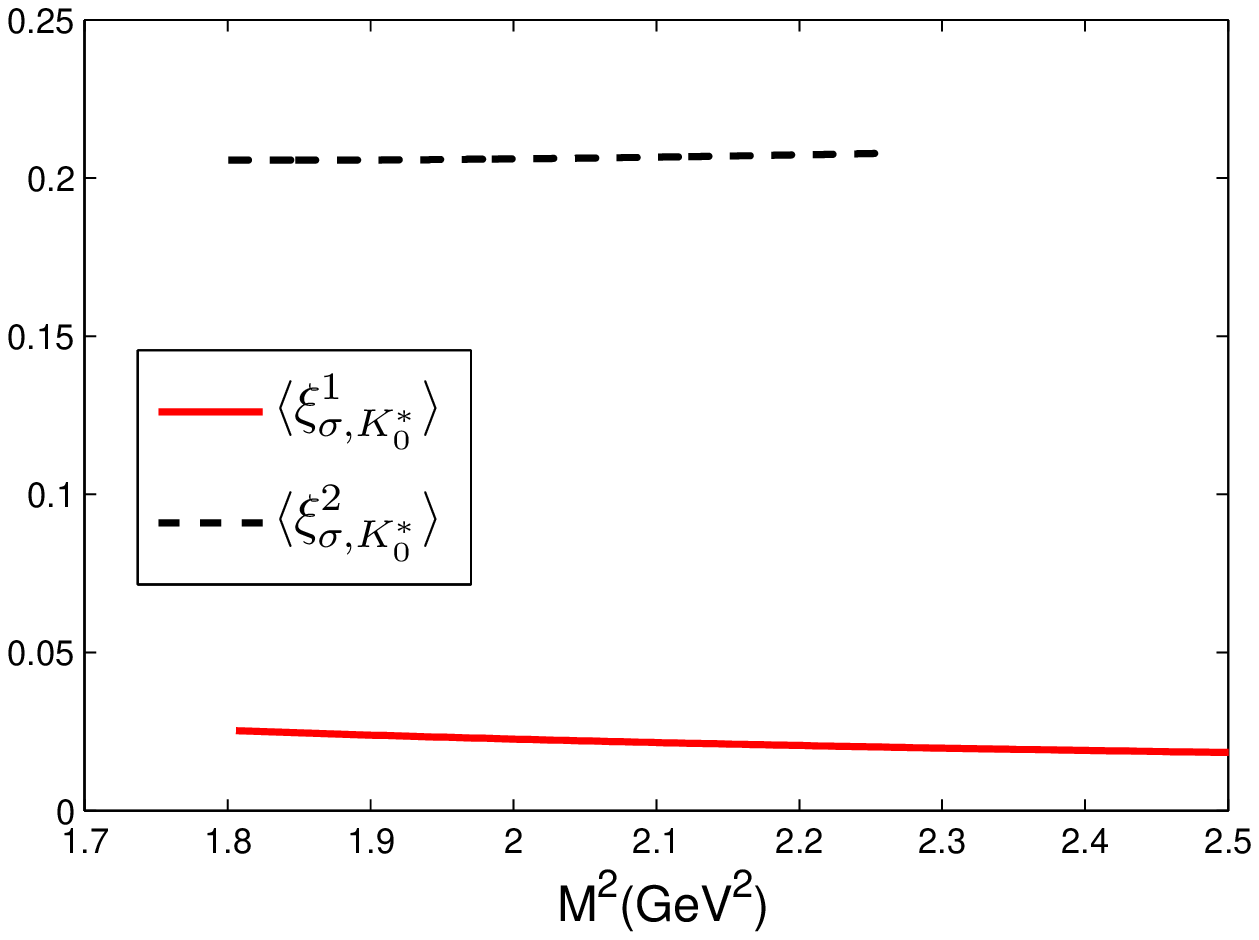}
\caption{Moments of scalar meson $K^{\ast}_0$ versus $M^2$ within their Borel windows.
}\label{figk0xi}
\end{figure}

\begin{figure}
\includegraphics[width=0.22\textwidth]{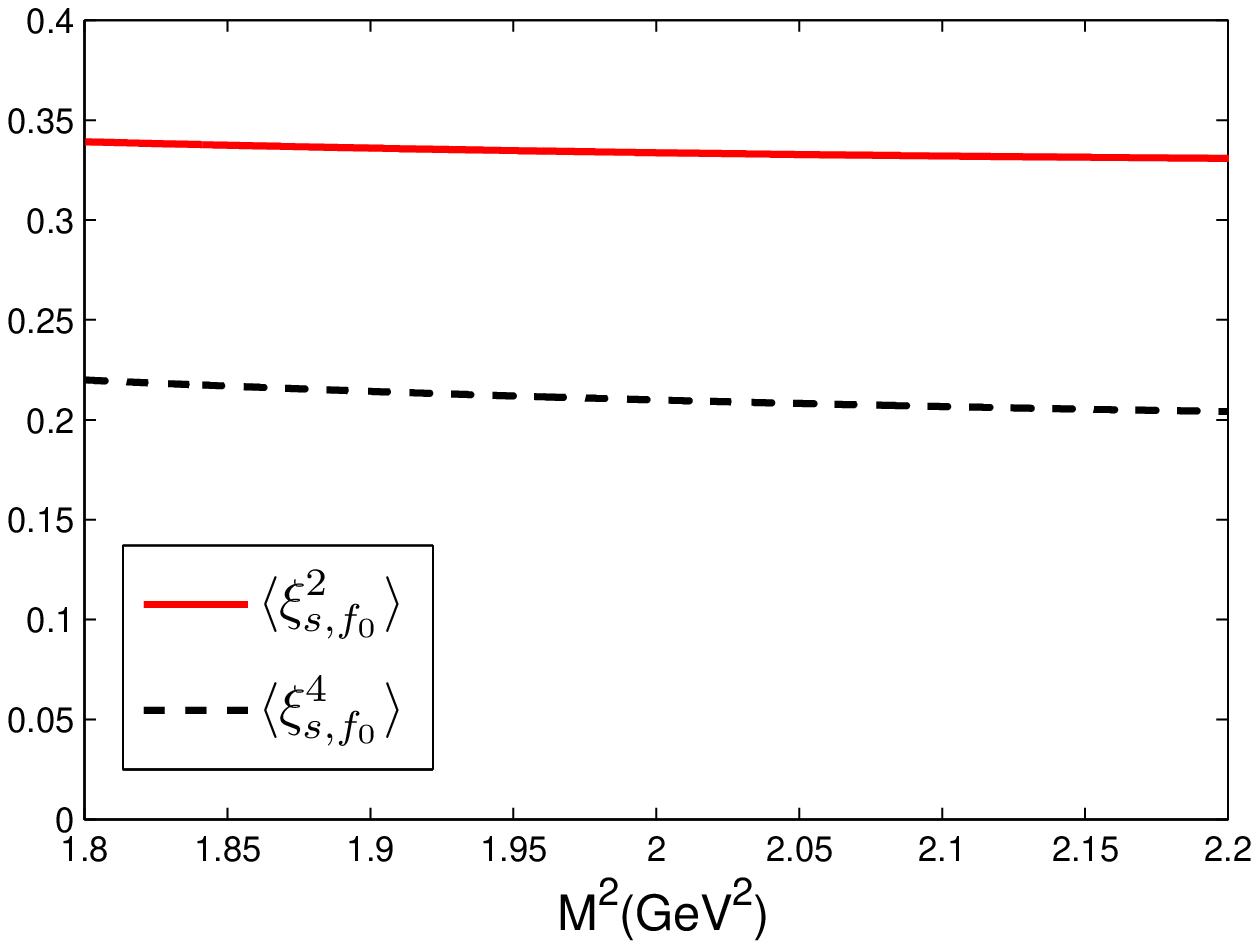}
\includegraphics[width=0.22\textwidth]{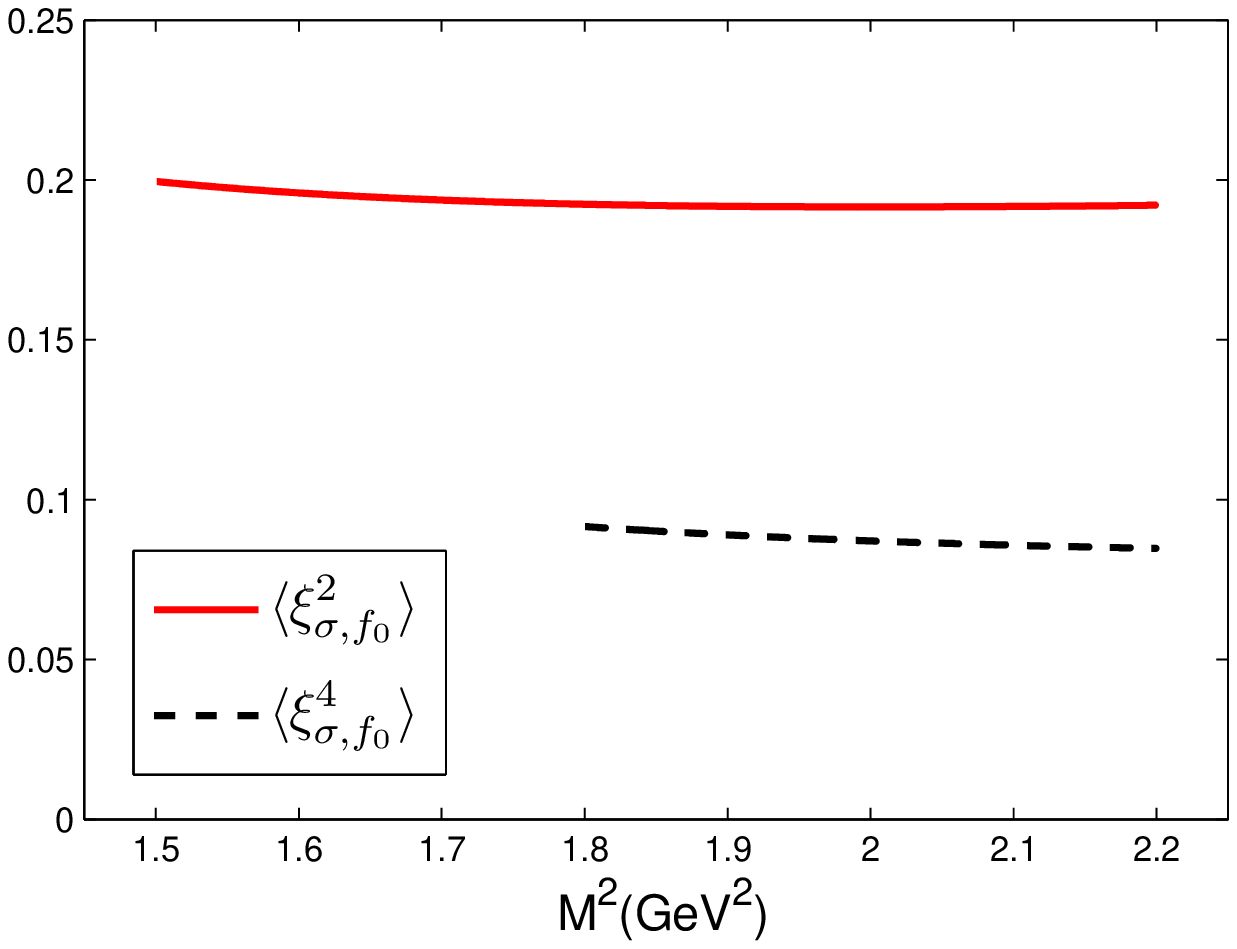}
\caption{Moments of scalar meson $f_0$ versus $M^2$ within their Borel windows.
}\label{figf0xi}
\end{figure}

By using the masses and decay constants for the scalar mesons, we further calculate the moments of these mesons from the sum rules (\ref{eq:resultslcdase}, \ref{eq:resultslcdate}, \ref{eq:resultslcdaso}, \ref{eq:resultslcdato}). Considering the conservation of charge parity and isospin symmetry, the odd moments for $a_0$ and $f_0$ meson twist-3 distribution amplitudes should vanish. While taking SU(3) symmetry breaking effects into account, the odd moments for $K^{\ast}_0$ meson are nonzero. Therefore, the moments we are aimed to derive are the second and fourth moments for $a_0$ and $f_0$ meson, and the first and second moments for $K^{\ast}_0$ meson. Similar to the way to get masses and decay constants, we first find the stable Borel window for the sum rules of each moments. The choice of Borel parameters for each meson, together with the moments for the scalar mesons, are collected in Tables \ref{tabmomentss} and \ref{tabmomentst}. In the Tables \ref{tabmomentss} and \ref{tabmomentst}, the Borel windows are determined by setting ${\rm SIX}<5\%$ and ${\rm CON}<30\%$. The moments for the scalar mesons versus $M^2$ are presented in Figs.(\ref{figa0xi}, \ref{figk0xi}, \ref{figf0xi}). It is noted that the moments within their Borel windows are very stable and their central values are
\begin{equation}
\begin{array}{ll}
\langle \xi_{s,a_0}^{2(4)} \rangle=0.369 \;(0.245), & \langle \xi_{\sigma,a_0}^{2(4)} \rangle=0.203 \;(0.093), \nonumber \\
 \langle \xi_{s,K^{\ast}_0}^{1(2)} \rangle =0.004\;(0.355), & \langle \xi_{\sigma,K^{\ast}_0}^{1(2)} \rangle =0.018\;(0.207), \nonumber \\
\langle \xi_{s,f_0}^{2(4)} \rangle=0.335 \;(0.212), &\langle \xi_{\sigma,f_0}^{2(4)} \rangle=0.196 \; (0.088),\nonumber
\end{array}
\end{equation}
It is noted that these moments are different from Ref.\cite{lv0612210}, especially for the moments of $K^{\ast}_0$ and $f_0$. By using the same input parameters, we can obtain consistent moments for $a_0$. Then, the differences are mainly caused by the different treatment of the mass terms (the $s$-quark mass terms) involved in the hard part calculation \footnote{Since the current $s$-quark mass is at the order of $\Lambda_{QCD}$, it is reasonable to assum that those terms will have sizable effects.}. Similar to the kaonic case~\cite{zhong}, it is found that those $m_s$-terms do provide sizable contributions. Thus, they should be treated consistently with those of the higher dimensional matrix elements. By taking all the mass terms consistently into consideration, a more reliable masses and decay constants, and hence more accurate moments can be obtained.

\subsection{A discussion on the scalar's three particle twist-3 LCDAs}

For each scalar meson $S$, there are three twist-3 distribution amplitudes $\phi^{s}_{S}$, $\phi^{\sigma}_{S}$ and $\phi_{3S}$. To be useful reference, we make a discussion on the three particle twist-3 LCDA, which, similar to the pseudoscalar and the vector cases \cite{vec3,phi3s1,phi3s2}, is defined as
\begin{widetext}
\begin{eqnarray}
&&\langle 0 | \bar{q_2}(x) \sigma_{\mu\nu} g G_{\alpha\beta}(-v x) q_{1}(-x)| S(q)\rangle = i \bar{f}_{3S} \big[q_\alpha(q_\mu\delta_{\nu\beta}-q_\nu\delta_{\mu\beta}) - (\alpha\leftrightarrow \beta)\big] \int {\cal D}\alpha_i\, e^{i qx(-\alpha_1+\alpha_2+v\alpha_3)} \phi_{3S}(\alpha_i)
\end{eqnarray}
\end{widetext}
where ${\cal D}\alpha_i  =d\alpha_1d\alpha_2d\alpha_3\delta(\alpha_1+\alpha_2+\alpha_3-1)$. Defining
\begin{displaymath}
R_1=\frac{1}{m_{a_0}}\frac{\bar{f}_{3a_0}}{\bar{f}_{a_0}} \;, \; R_2=\frac{1}{m_{K^*_0}}\frac{\bar{f}_{3K^*_0}}{\bar{f}_{K^*_0}} \;\,\; R_3=\frac{1}{m_{f_0}} \frac{\bar{f}_{3f_0}}{\bar{f}_{f_0}},
\end{displaymath}
by using the recurrence relations for the moments among the twist-3 LCDAs that are derived by using the equations of motion and the conformal symmetry~\cite{phi3s1,phi3s2,phi3s3} \footnote{As an estimation, we neglect the meson mass effects in deriving those relations.}, we obtain
\begin{eqnarray}
\langle\xi^2_{\sigma,a_0}\rangle &=& \frac{1}{5} \langle\xi^0_{\sigma,a_0}\rangle +\frac{12}{5}R_{1}-\frac{8}{5}R_{1}\langle\langle\alpha_3\rangle\rangle_{a_0} \\
\langle\xi^2_{s,a_0}\rangle &=& \frac{1}{3}\langle\xi^0_{s,a_0}\rangle +4R_1 \\
\langle\xi^2_{\sigma,K^*_0}\rangle &=& \frac{3}{5} \langle\xi^2_{s,K^*_0}\rangle -\frac{8}{15}R_{2}\langle\langle\alpha_3\rangle\rangle_{K^*_0} \\
\langle\xi^2_{s,K^*_0}\rangle &=& \frac{1}{3}\langle\xi^0_{\sigma,K^*_0}\rangle +4R_2 \\
\langle\xi^2_{\sigma,f_0}\rangle &=& \frac{1}{5} \langle\xi^0_{\sigma,f_0}\rangle +\frac{12}{5}R_{3}-\frac{8}{5}R_{3}\langle\langle\alpha_3\rangle\rangle_{f_0} \\
\langle\xi^2_{s,f_0}\rangle &=& \frac{1}{3}\langle\xi^0_{s,f_0}\rangle +4R_3
\end{eqnarray}
where $\langle \!\langle(\alpha_2 -\alpha_1+v\alpha_3)^n\rangle \!\rangle
=\int {\cal D}\alpha_{i}\phi_{3 S}(\alpha_i)(\alpha_2 -\alpha_1+v \alpha_3)^n$ defines the moments of the three-particle twist-3 LCDAs. Using the central values for the two particle twist-3 LCDA moments, we obtain
\begin{eqnarray}
&&\langle\langle\alpha_3\rangle\rangle_{a_0}\sim 1.29 \;\;{\rm and}\;\; f_{3a_0} \sim 4.82\times10^{-3} {\rm GeV}^2 \ , \\
&&\langle\langle\alpha_3\rangle\rangle_{K^*_0}\sim 2.08 \;\;{\rm and}\;\; f_{3K^*_0} \sim 2.76\times10^{-3} {\rm GeV}^2 \ , \\
&&\langle\langle\alpha_3\rangle\rangle_{f_0}\sim 7.50 \;\;{\rm and}\;\; f_{3a_0} \sim 2.56\times10^{-4} {\rm GeV}^2 \ .
\end{eqnarray}
These equations show that all $R_i$ are less than $1\%$, so in usual calculations, one can safely neglect the scalar meson's three-particle twist-3 LCDA in comparison to the two-particle twist-3 LCDAs. This is different from the pion and the kaon, whose similar ratios $R_{\pi}$ ($R_{K}$) $\sim 5\%$~\cite{phi3s3}; then for the pionic or kaonic processes, one may need to take $\phi_{3\pi}$ or $\phi_{3K}$ into consideration.

\subsection{Scalar meson distribution amplitudes}

The scalar mesons' twist-3 LCDAs can be expanded into a series of Gegenbauer polynomials \cite{Braun:ppnp51, Chernyak:pr112}
\begin{eqnarray}
\phi_{S}^{s}(u,\mu)&=&1+\sum_{m=1}^{\infty }
a_{m}(\mu)C_{m}^{1/2}(2u-1), \label{eq:GegenA} \\
\phi_{S}^{\sigma}(u,\mu)&=&6u \bar{u}[1+\sum_{m=1}^{\infty}b_{m}(\mu)
C_{m}^{3/2}(2u-1)].   \label{eq:GegenB}
\end{eqnarray}
where $C_{m}^{3/2,1/2}(2u-1)$ are Gegenbauer polynomials. The Gegenbauer moments $a_m$, $b_m$ can be related to moments $\langle \xi^m_s \rangle$ and $\langle \xi^m_{\sigma} \rangle$ defined in Eqs.(\ref{eq:xidf0}, \ref{eq:xidf1}). By using the orthogonality of Gegenbauer polynomials
\begin{eqnarray}
&&\int^1_0  du C_n^{1/2}(2u-1)C_m^{1/2}(2u-1)=\frac{1}{2 n+1}\delta_{m n}, \\
&&\int^1_0  du u(1-u)C_n^{3/2}(2u-1)C_m^{3/2}(2u-1)= \nonumber\\
&& \frac{(n+2)(n+1)}{4 (2n+3)}\delta_{m n} ,
\end{eqnarray}
we obtain the usual Gegenbauer moments:
\begin{eqnarray}
&& a_1=3 \langle \xi_1\rangle, \; a_2=\frac{5}{ 2} \left( 3 \langle \xi_2\rangle-1 \right),\nonumber\\
&& a_4=\frac{9}{ 8} \left( 35 \langle \xi_4 \rangle-30 \langle \xi_2 \rangle+3 \right),\label{relationA}\\
&& b_1=\frac{5}{3} \langle \xi_1 \rangle, \; b_2=\frac{7 }{ 12} \left( 5
\langle \xi_2\rangle -1 \right), \nonumber\\
&& b_4=\frac{11}{ 24} \left(21 \langle \xi_4\rangle -14 \langle \xi_2 \rangle+1 \right) .\label{relationB}
\end{eqnarray}

\begin{widetext}
\begin{center}
\begin{table*}[htb]
\caption{Gegenbauer moments for $\phi^{s}_{S}$ at two energy scales $1$ GeV and $2.4$ GeV, where $S$ stands for the scalar meson $a_0$, $k_0^{\ast}$ or $f_0$ respectively.}\label{tabgenmomentss}
\begin{tabular}{|c|c|c|c|c|c|c|}
\hline
   &\multicolumn{3}{c|}{$\mu=1\rm{GeV}$} &\multicolumn{3}{c|}{$\mu=2.4\rm{GeV}$} \\
\cline{2-7}
\raisebox{2.0ex}[0pt]{mesons}
   &$a_1$ &$a_2$  &$a_4$ &$a_1$ &$a_2$  &$a_4$ \\
\hline
$a_0$   &0 &$0.302\sim0.323$  &$0.491\sim0.944$  &0 &$0.225\sim0.241$  &$0.307\sim0.591$\\
\hline
$k_0^{\ast}$ &$0.0109\sim0.0143$  &$0.163\sim 0.211$ & $\sim$ &$0.0095\sim0.0125$  &$0.122\sim 0.157$ & $\sim$ \\
\hline
$f_0$   &0   &$-0.022\sim 0.049$  &$0.140 \sim 0.892$  &0   &$-0.016\sim 0.036$  &$0.088 \sim 0.56$\\
\hline
\end{tabular}
\end{table*}
\end{center}
\end{widetext}

\begin{widetext}
\begin{center}
\begin{table*}[htb]
\caption{Gegenbauer moments for $\phi^{\sigma}_{S}$ at two energy scales $1$ GeV and $2.4$ GeV, where $S$ stands for the scalar meson $a_0$, $k_0^{\ast}$ or $f_0$ respectively.}
\begin{tabular}{|c|c|c|c|c|c|c|}
\hline\hline
   &\multicolumn{3}{c|}{$\mu=1$ GeV} &\multicolumn{3}{c|}{$\mu=2.4$ GeV} \\
\cline{2-7}
\raisebox{2.0ex}[0pt]{meson}
   &$b_1$ &$b_2$  &$b_4$ &$b_1$ &$b_2$  &$b_4$ \\
\hline
$a_0$   &0 &$0.011\sim0.013$  &$0.050 \sim 0.073 $  &0 &$0.008\sim0.009$  &$0.031 \sim 0.045 $ \\
\hline
$k_0^{\ast}$ &$0.0216\sim0.0468$  &$0.019\sim 0.028$   & $\sim$ &$0.017\sim0.037$  &$0.013\sim 0.020$ & $\sim$ \\
\hline
$f_0$   &0   &$-0.0291\sim -0.0014$  &$0.030\sim0.106$  &0  &$-0.0206\sim -0.0010$  &$0.018\sim 0.065$\\
\hline
\end{tabular}
\label{tabgenmomentst}
\end{table*}
\end{center}
\end{widetext}

Using the moments $\langle \xi^n_{s} \rangle$ and $\langle \xi^n_{\sigma} \rangle$ derived in the last subsection, one can obtain the Gegenbauer moments $a_m$ and $b_m$ by using the relations (\ref{relationA},\ref{relationB}). For convenience, we present the first two non-zero ones for each meson at two different energy scales in Tables \ref{tabgenmomentss} and \ref{tabgenmomentst}, where the Gegenbauer moments at $\mu=2.4$ GeV are obtained by using the evolution equations.

\begin{figure}[htb]
\includegraphics[width=0.23\textwidth]{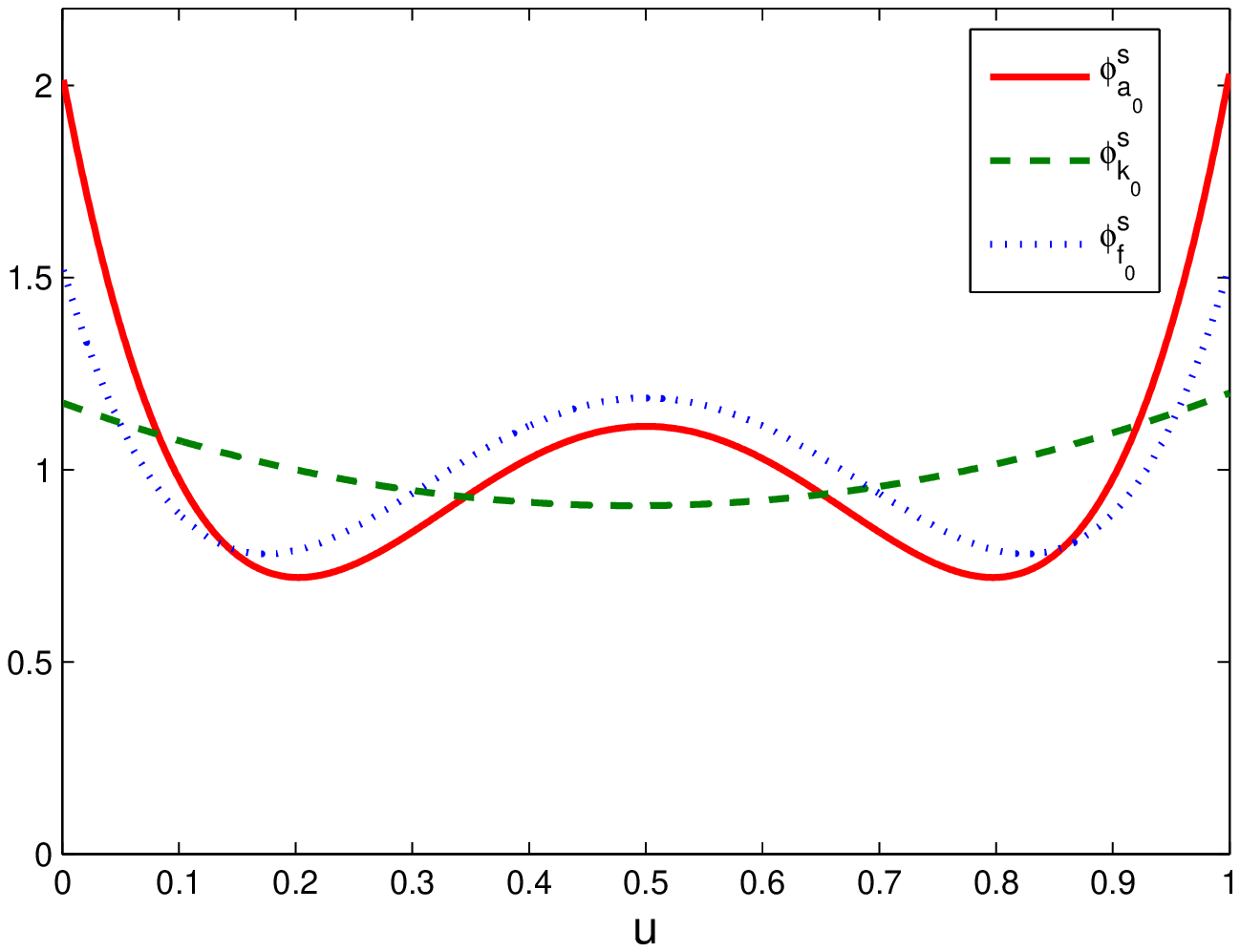}
\includegraphics[width=0.23\textwidth]{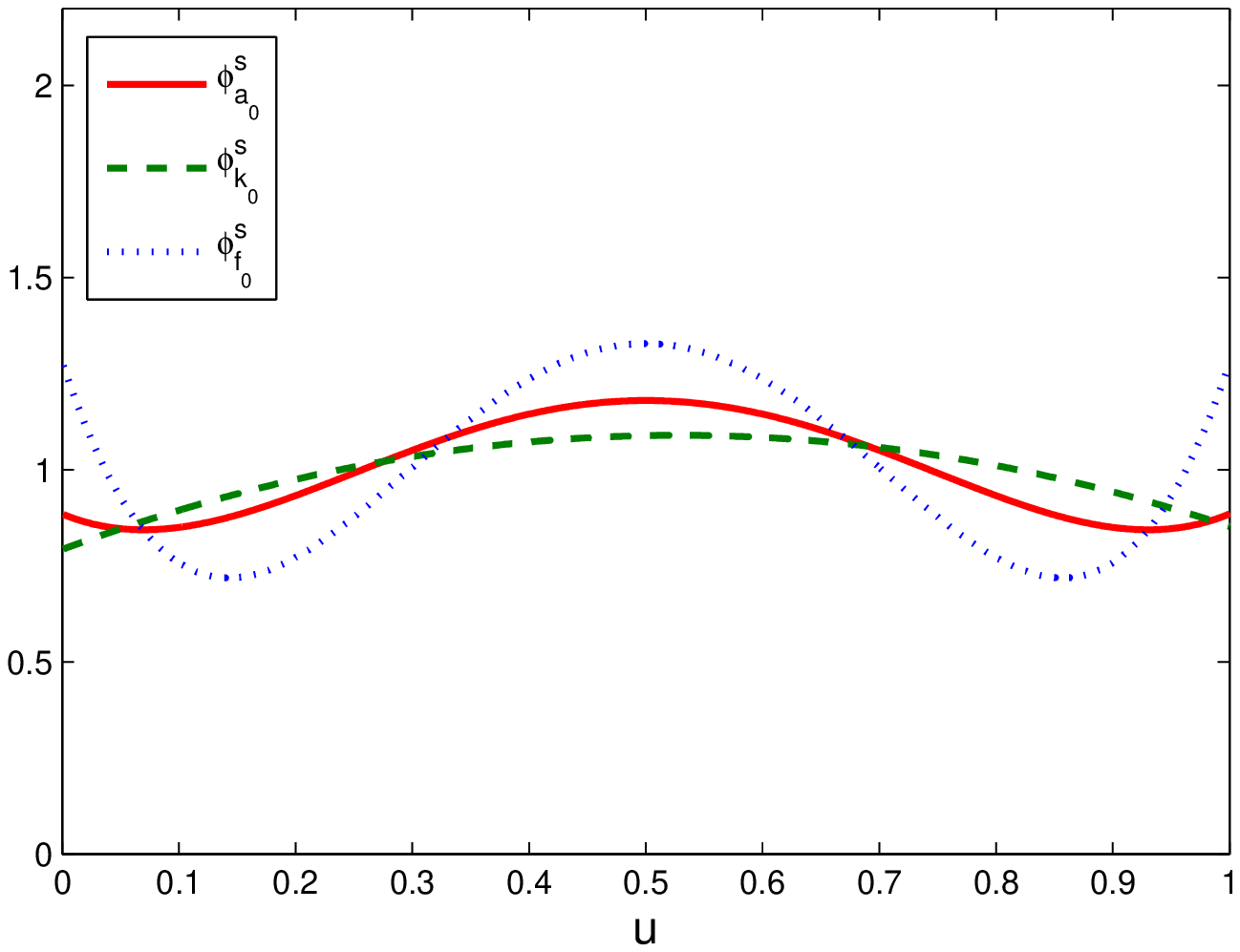}
\caption{Twist-3 LCDAs $\phi_S^s(u)$ for scalar mesons at the scale $\mu=1$ GeV, where the solid, the dashed and the dotted lines denote the LCDAs of $a_0$, $K_0^{\ast}$ and $f_0$ respectively. The left diagram is for our LCDAs and the right one is for Ref.\cite{lv0612210}. }
\label{figtwist3dass}
\end{figure}

\begin{figure}[htb]
\includegraphics[width=0.23\textwidth]{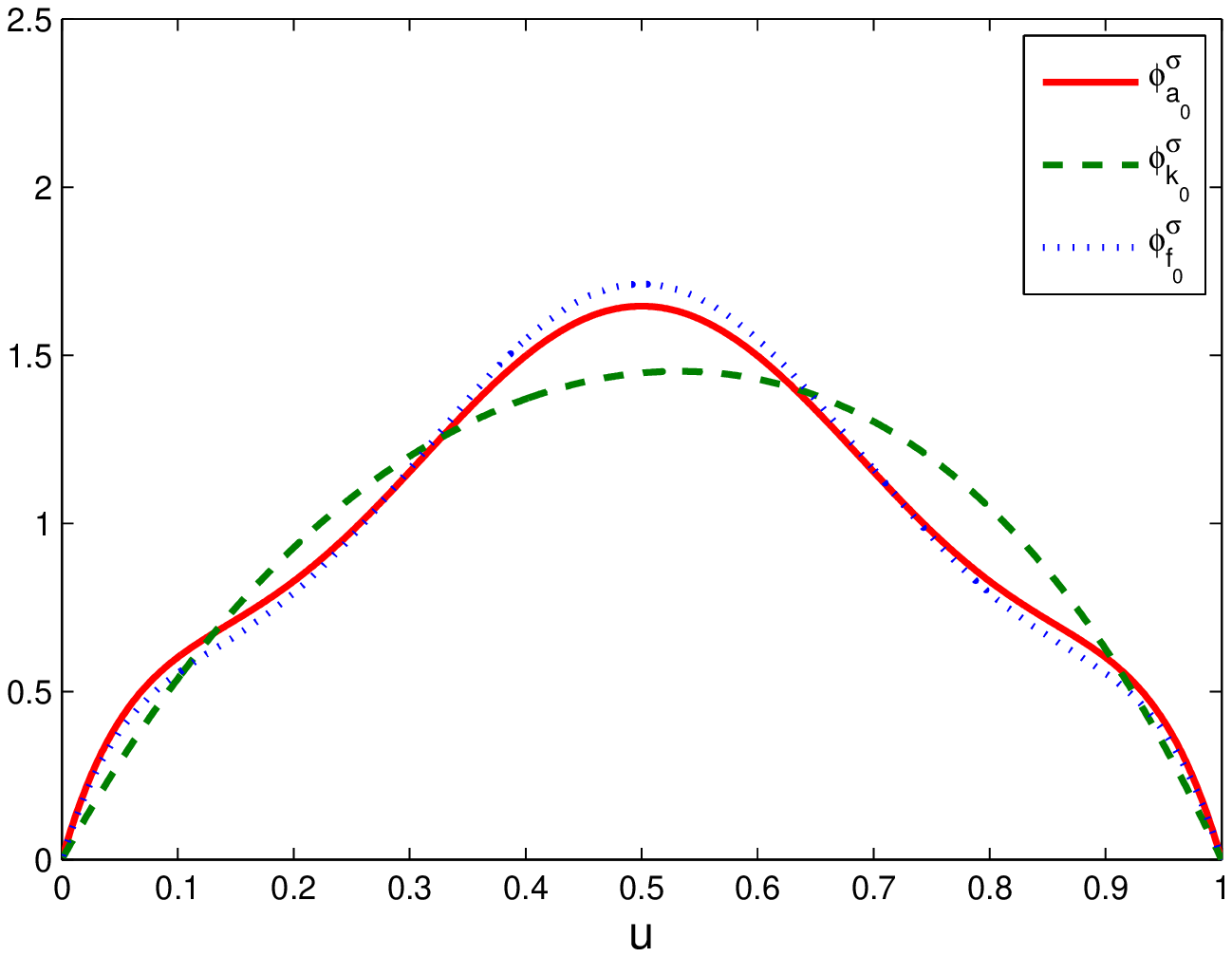}
\includegraphics[width=0.23\textwidth]{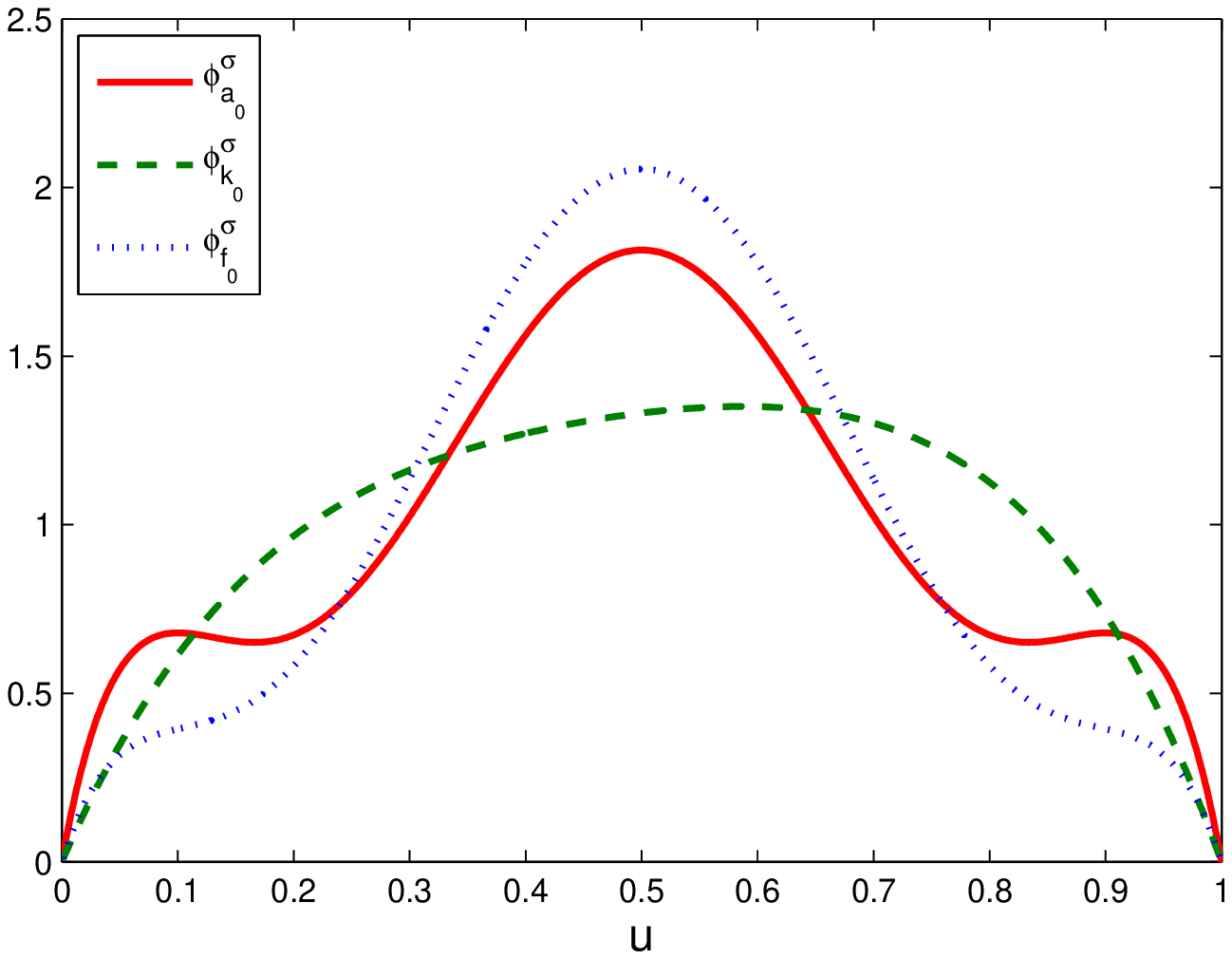}
\caption{Twist-3 LCDAs $\phi_S^{\sigma}(u)$ for scalar mesons at the scale $\mu=1$ GeV, where the solid, the dashed and the dotted lines denote the LCDAs of $a_0$, $K_0^{\ast}$ and $f_0$ respectively. The left diagram is for our LCDAs and the right one is for Ref.\cite{lv0612210}. }
\label{figtwist3dast}
\end{figure}

The LCDAs $\phi_S^s(u)$ and $\phi_S^{\sigma}(u)$ at $\mu=1$ GeV are presented in Figs.(\ref{figtwist3dass},\ref{figtwist3dast}). As a comparison we also present the results of Ref.\cite{lv0612210} in Figs.(\ref{figtwist3dass},\ref{figtwist3dast}).

\subsection{Properties of the $B \to S$ transition form factors}

\begin{figure}[htb]
\includegraphics[width=0.35\textwidth]{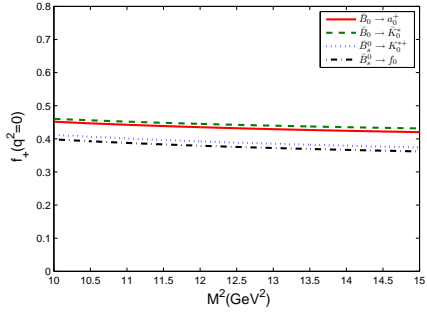}
\caption{Prediction of $f_+(q^2=0)$ within the LCSR approach with chiral currents versus the Borel parameter $M$. The solid, the dashed, the dotted and the dash-dot lines are for $\bar{B}_0 \to a_0^+(1450)$, $\bar{B}_0 \to \bar{K}^*_0(1430)$, $\bar{B}_s^0 \to K_0^{*+}(1430)$ and $\bar{B}_s^0 \to f_0(1500)$ respectively. The input parameters are taken as their central values. }\label{figbsftw3fp}
\end{figure}

In doing the numerical calculation we adopt \cite{pdg,cheng2006,Wang:prd78}: $m_{B_0}=5.279$ GeV, $m_{B_s}=5.368$ GeV, $f_{B_0}=(0.19\pm 0.02)$ GeV, $f_{B_s}=(0.23\pm 0.02)$ GeV, $m_b=(4.68\pm 0.03)$ GeV. As for the energy scale of the $B \to S$ transitions, we adopt $\mu_b=\sqrt{m_{B_s}^2-m_b^2} \sim 2.4$ GeV. To calculate the form factors one has to evolve the parameters to the scale $\mu_b$, which can be derived by using the renormalization group equations (\ref{RGE},\ref{momentsRGE}). We take the threshold parameter $S^{B_0}_0=S^{B_s}_0=(33 \pm 1)$ ${\rm GeV}^2$~\cite{Sun:prd83}. The Borel window is determined by requiring both the contributions from higher excited resonances and continuum states be less than 30\%, which results in $10\rm{GeV} ^2 \leq M^2 \leq 15 \rm{GeV} ^2$. As shown in Fig.(\ref{figbsftw3fp}), the sum rules for the form factors at the large recoil region vary slightly within such Borel window.

\begin{table}[htb]
\caption{Numerical results for the $B\to S$ transition form factors at the large recoil point $q^2=0$, where $S$ stands for the scalar mesons $a^+$, $f_0$, $K^{*+}_0$ and $\bar{K}^{*}_0$, respectively. }
\begin{center}
\begin{tabular}{lcccc}
\hline \hline
processes &Methods &$f_+$ &$f_-$ &$f_T$  \\ \hline
$\bar{B}^0 \to a^{+}_0(1450)$ &This work &$0.44^{+0.06}_{-0.05}$ &$-0.26^{+0.06}_{-0.05}$ &$0.43^{+0.06}_{-0.05}$ \\
	&LCSR \cite{Wang:prd78} &$0.52$   &$-0.44$  &$0.66$ \\
	&LCSR \cite{Sun:prd83}  &$0.53$   &$-0.53$  &$\sim$ \\
	&pQCD \cite{Li:prd79} &$0.68$     &  $\sim$  &$0.92$ \\
	&CLF \cite{Cheng:prd69} &$0.26$   &   $\sim$ &$\sim$ \\
\hline
$\bar{B}^0_s \to f_0(1500)$   &This work   &$0.38^{+0.04}_{-0.04}$ &$-0.24^{+0.04}_{-0.04}$ &$0.40^{+0.04}_{-0.04}$ \\
	&LCSR \cite{Wang:prd78} &$0.43$   &$-0.37$  &$0.56$ \\
        &LCSR \cite{Sun:prd83}  &$0.41$   &$-0.41$  &$0.59$ \\
	&pQCD \cite{Li:prd79} &$0.60$     &   $\sim$   &$0.82$\\
\hline
$\bar{B}^0_s \to K^{*+}_0(1430)  $ &This work &$0.39^{+0.04}_{-0.04}$ &$-0.25^{+0.05}_{-0.05}$ &$0.41^{+0.04}_{-0.04}$ \\
	&LCSR \cite{Wang:prd78} &$0.42$   &$-0.34$  &$0.52$ \\
	&LCSR \cite{Sun:prd83}  &$0.44$   &$-0.44$  &$\sim$\\
	&SR   \cite{Yang:prd73} &$0.24$   &   $\sim$    &$\sim$\\
	&pQCD \cite{Li:prd79} &$0.56$     &   $\sim$   &$0.72$\\
\hline
$\bar{B}^0 \to \bar{K^*_0}(1430) $ &This work  &$0.45^{+0.06}_{-0.05}$ &$-0.28^{+0.06}_{-0.06}$ &$0.46^{+0.06}_{-0.05}$ \\
	&LCSR \cite{Wang:prd78} &$0.49$   &$-0.41$  &$0.60$ \\
        &LCSR \cite{Sun:prd83}  &$0.49$   &$-0.49$  &$0.69$ \\
	&SR   \cite{Aliev:prd76} &$0.31$  &$-0.31$  &$-0.26$ \\
	&pQCD \cite{Li:prd79}   &$0.60$   &  $\sim$   &$0.78$\\
	&CLF \cite{Cheng:prd69} &$0.26$   &  $\sim$   & $\sim$\\
	&LFQM \cite{Chen:prd75} &$-0.26$ &$0.21$    &$-0.34$ \\
\hline
\end{tabular}
\end{center}\label{tabbsftw3}
\end{table}

\begin{figure}[htb]
\includegraphics[width=0.35\textwidth]{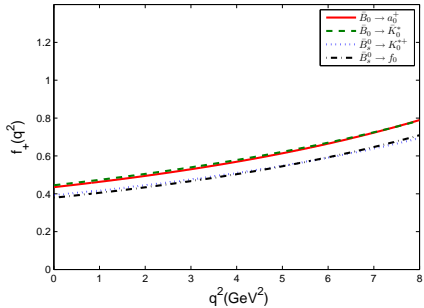}
\caption{Form factors $f_+(q^2)$ within the LCSR approach with chiral currents. The solid, the dashed, the dotted and the dash-dot lines are for $\bar{B}_0 \to a_0^+(1450)$, $\bar{B}_0 \to \bar{K}^*_0(1430)$, $\bar{B}_s^0 \to K_0^{*+}(1430)$ and $\bar{B}_s^0 \to f_0(1500)$ respectively. Here the Borel parameter $M^2=12$ ${\rm GeV}^2$ within the LCSR approach. }\label{figbsffpq}
\end{figure}

We present the $B \to S$ form factors at the large recoil point $q^2=0$ in Table \ref{tabbsftw3}, where the theoretical uncertainties are around $20 \%$, which are caused by varying the Borel parameter $M$, the threshold values $s^{B_0}_0$ and $s^{B_s}_0$, the $b$ quark mass, the decay constants of the scalar mesons, and the Gegenbauer moments for the twist-3 LCDAs of the scalar mesons, within their reasonable regions. As a comparison, we also put several typical estimations from the SR, the LCSR and the pQCD approaches \cite{Wang:prd78,Sun:prd83,Li:prd79,Cheng:prd69,Cheng:prd69,Yang:prd73,Aliev:prd76,Chen:prd75}, in Table \ref{tabbsftw3}. The form factors $f_+(q^2)$ for various scalar mesons $a_0^+(1450)$, $\bar{K}^*_0(1430)$, $K_0^{*+}(1430)$ and $f_0(1500)$ rise slightly with the increment of $q^2$, as can be shown explicitly in Fig.(\ref{figbsffpq}).

\subsection{Two semi-leptonic channels $B \to S l\bar{\nu}_{l}$ and $B \to S l\nu_{l}$}

\begin{figure}
\includegraphics[width=0.23\textwidth]{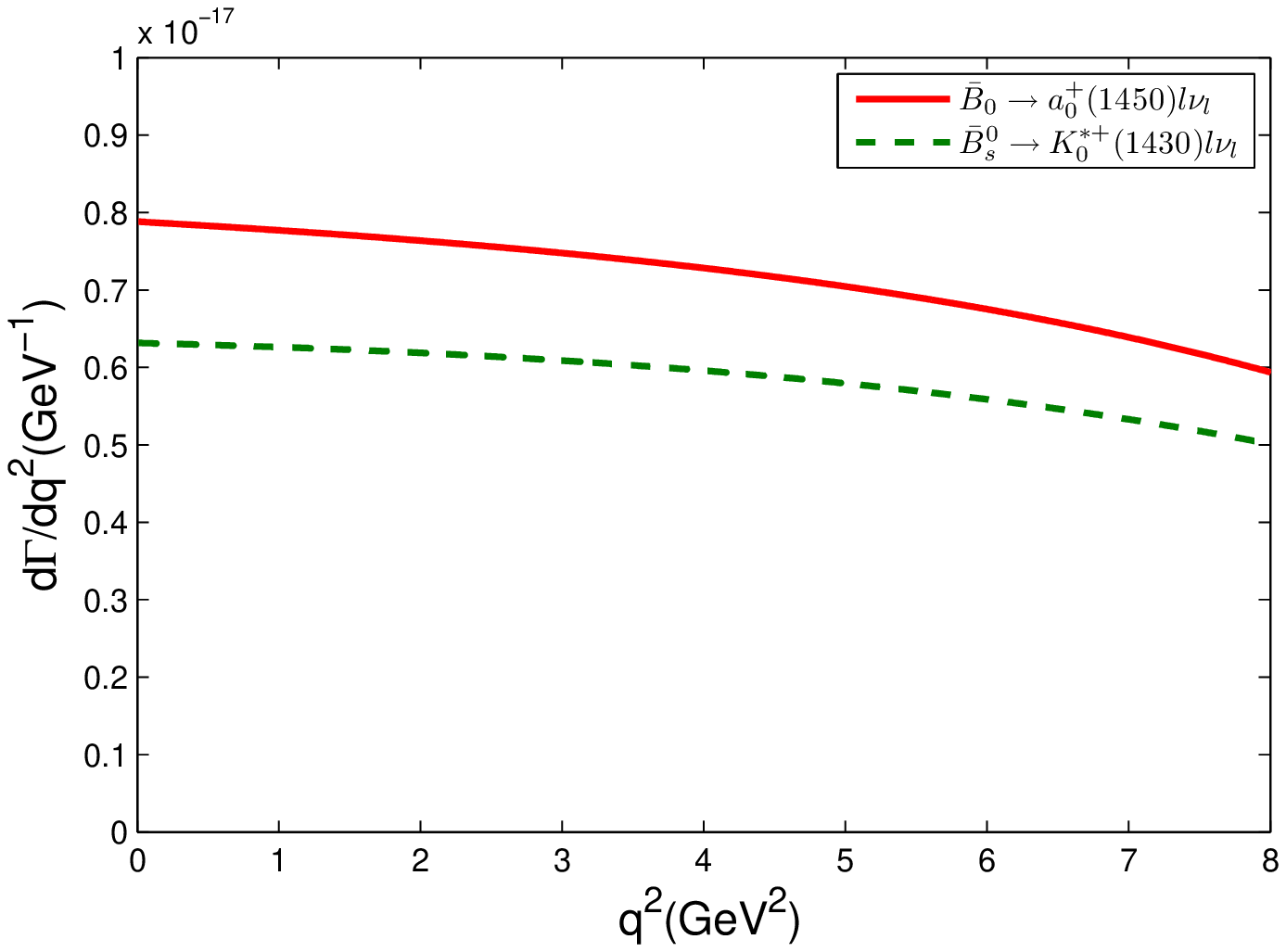}
\includegraphics[width=0.23\textwidth]{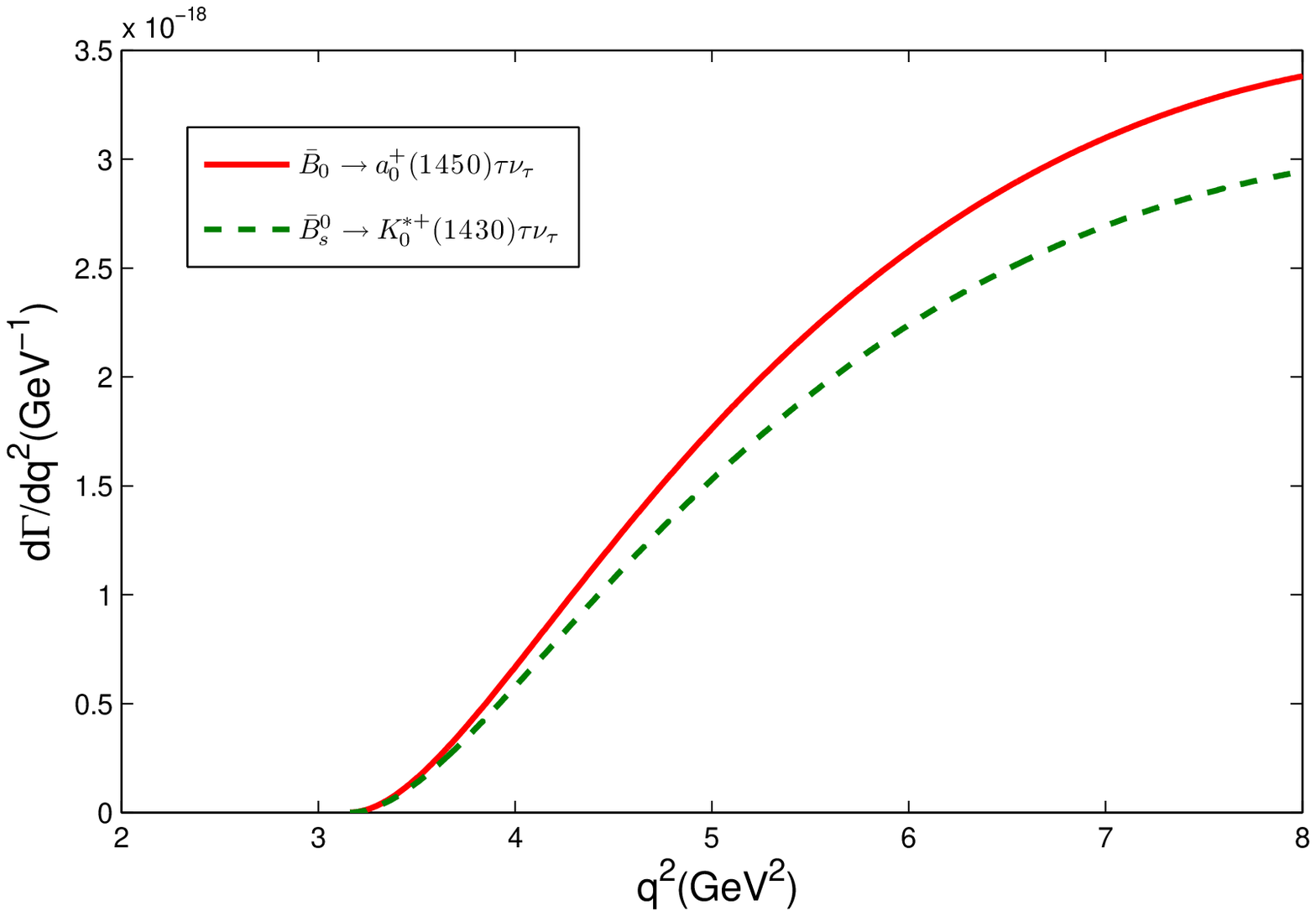}
\caption{Differential decay width for the $B \to S l \bar{\nu}_l$ decays. Here the massless lepton $l=e$ or $\mu$. Here $S$ stands for $a^+_0(1450)$ or $K^{+\ast}_0(1430)$. }\label{bslnul}
\end{figure}

\begin{figure}
\includegraphics[width=0.23\textwidth]{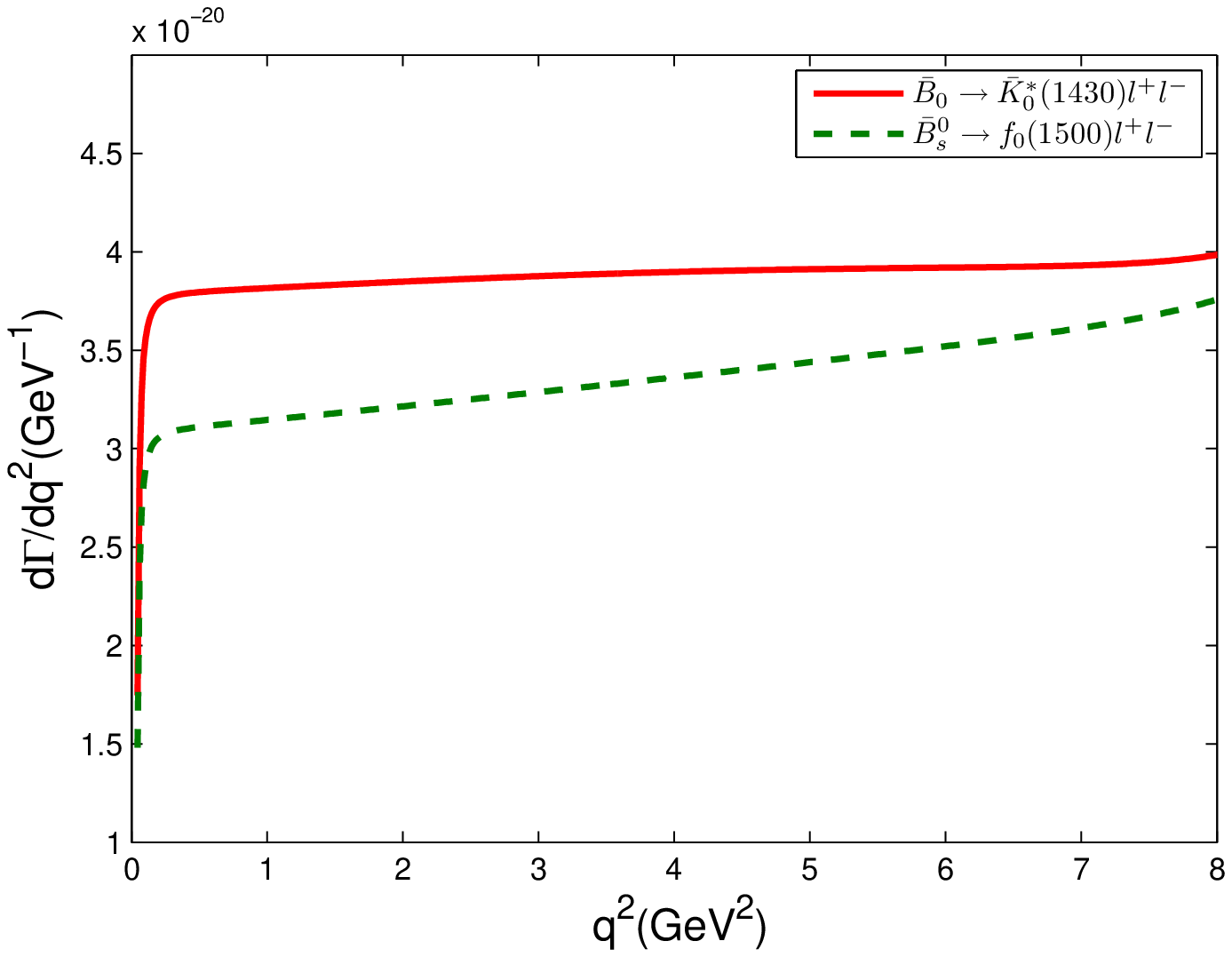}
\includegraphics[width=0.23\textwidth]{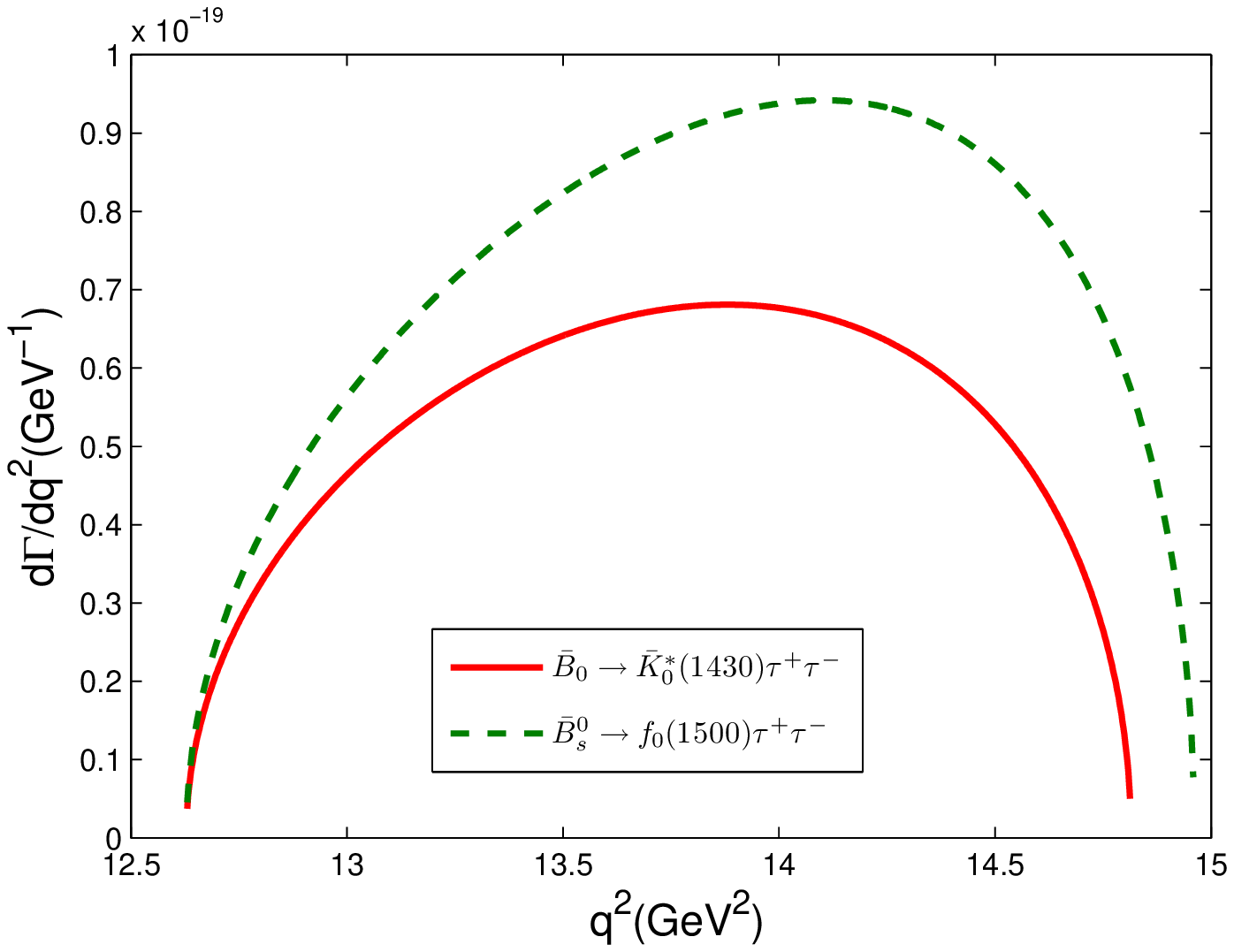}
\caption{Differential decay width for the rare $B \to S l \bar{l}$ decays. Here the massless lepton $l=e$ or $\mu$. Here $S$ stands for $\bar{K}^{\ast}_0(1430)$, $f_0(1500)$. }\label{bsll}
\end{figure}

Next, as an application of the $B\to S$ transition form factors, we discuss the properties of the two semi-leptonic channels, $B \to S l\bar{\nu}_{l}$ and $B \to S l\nu_{l}$, whose differential decay widths can be expressed as
\begin{eqnarray}
&&\frac{d\Gamma}{dq^2}(B_{(s)}\to S l \bar{\nu}_l)=\frac{G_F^2|V_{ub}|^2}{192 \pi^3 m_B^3}\frac{q^2-m_l^2}{(q^2)^2} \nonumber\\
&& \sqrt{ \frac{(q^2-m_l^2)^2}{q^2}}\sqrt{\frac{(m_B^2-m_S^2-q^2)^2} {4q^2}-m_S^2}\bigg[(m_l^2+2q^2)\nonumber\\
&&(q^2-(m_B-m_S)^2)(q^2-(m_B+m_S)^2)f_+^2(q^2)+ \nonumber\\
&& 3m_l^2(m_B^2-m_S^2)^2\left(f_+(q^2)+ \frac{q^2} {m_B^2-m_S^2}f_-(q^2)\right)^2\bigg],\label{eq:widthsemi}
\end{eqnarray}
and
\begin{eqnarray}
&&\frac{d\Gamma }{dq^2}(B_{(s)}\to S l \bar{l})=\frac{G_{F}^{2}\left| V_{tb}V_{ts}\right| ^{2}m_{B}^{3}\alpha _{em}^{2}}{1536 \pi ^{5}} \nonumber\\
&& \quad\left( 1-\frac{4r_{l}}{s}\right) ^{1/2}\left[ \left( 1+\frac{2r_{l}}{s}\right)\varphi _{S}^{3/2} \alpha
_{S}+\varphi _{S}^{1/2}r_{l}\delta _{S}\right], \label{eq:widthrare}
\end{eqnarray}
where $m_l$ denotes the mass of a final state lepton, and
\begin{eqnarray}
s &=&q^{2}/m_{B}^{2}, \; r_{l}=m_{l}^{2}/m_{B}^{2},\; r_{S}=m_{S}^{2}/m_{B}^{2}, \\
\varphi _{S} &=&\left( 1-r_{S}\right) ^{2}-2s\left( 1+r_{S}\right) +s^{2} , \\
\alpha _{S} &=& \left| C_{9}^{eff}{f_{+}\left( q^{2}\right) }
-2\frac{C_{7}f_{T}\left( q^{2}\right) }{1+\sqrt{r_{S}}}\right|
^{2}+\left|C_{10}{f_{+}\left( q^{2}\right) }\right| ^{2} , \\
\delta _{S} &=& 6\left| C_{10}\right| ^{2} \Big\{ \left[ 2\left(
1+r_{S}\right) -s\right] \left| {f_{+}\left( q^{2}\right)} \right|
^{2} + \nonumber\\
&& \left( 1-r_{S}\right) 2{\rm{Re}} [ f_{+}( q^{2})
f^{\ast}_{-}(q^2) ] +s\left| f_{-}(q^2)\right| ^{2}\Big\}.
\end{eqnarray}
Except for the $B\to S$ transition form factors, we adopt the same input parameters as those of Ref.\cite{Sun:prd83}. Our results for the curves for the differential decay rates are presented in Figs.(\ref{bslnul}, \ref{bsll}), which have a different arising trends versus the increment of $q^2$ in comparison to that of Ref.\cite{Sun:prd83}. In Ref.\cite{Sun:prd83}, the authors also adopt the chiral currents in calculating the $B\to S$ transition form factors. However their chiral currents are different from ours, where only the twist-2 scalar LCDAs are remained and the twist-3 scalar LCDAs make no contributions. Those two different QCD sum rules can be taken as a cross check of the chiral LCSR approach and be confirmed by the future experiments.

\section{Summary}

The masses, the decay constants and the twist-3 LCDAs of the scalar mesons $a_0$, $K^{\ast}_0$ and $f_0$ have been investigated in this work by using the QCD sum rules together with the QCD background field theory. Our present estimations are improved by a consistent treatment of the mass effects and a RG improved treatment of the input parameters. It is found that the $m_s$-terms provide sizable contributions, which should be treated consistently with those of higher dimensional matrix elements. While, by taking all the mass terms consistently into consideration, a more reliable masses and decay constants, and hence more accurate moments can be obtained.

As for $a_0$ meson, the second and forth moments of the twist-3 LCDAs $\phi^{s,\sigma}_{a_0}$ are
\begin{displaymath}
\langle \xi_{s,a_0}^{2(4)} \rangle=0.369 \;(0.245)\;{\rm and}\; \langle \xi_{\sigma,a_0}^{2(4)} \rangle=0.203 \;(0.093),
\end{displaymath}
whose uncertainties are about $10\%$; as for the $K^*_0$ meson, the first and second moments of the twist-3 LCDAs $\phi^{s,\sigma}_{K^{\ast}_0}$ are
\begin{displaymath}
\langle \xi_{s,K^{\ast}_0}^{1(2)} \rangle =0.004\;(0.355)\;{\rm and}\; \langle \xi_{\sigma,K^{\ast}_0}^{1(2)} \rangle =0.018\;(0.207),
\end{displaymath}
whose uncertainties are about $10\%- 15\%$; as for the case of $f_0$ meson, the second and forth moments of the twist-3 LCDAs $\phi_{K^*_0}^{s,\sigma}$ are
\begin{displaymath}
\langle \xi_{s,f_0}^{2(4)} \rangle=0.335 \;(0.212)\;{\rm and}\; \langle \xi_{\sigma,f_0}^{2(4)} \rangle=0.196 \; (0.088),
\end{displaymath}
whose uncertainties are within $15\%$.

As an application of these twist-3 LCDAs, we have studied the $B \to S$ transition form factors and their corresponding decay rates. For the purpose, the chiral currents are adopted in our LCSR calculation of $B\to S$ transition form factors, in which the twist-3 LCDAs make dominant contributions. Our results for the transition form factors at the large recoil point $q^2 \simeq 0$ are consistent with those obtained in the literature; while, the arising trends for the form factors versus $q^2$, and hence the differential decay widths for the $B\to S$ semileptonic decays, are somewhat different. This can be checked by a future experimental data.

Our results for the twist-3 LCDAs are useful for pQCD calculation. Basing on the Gegenbauer moments of the twist-3 LCDAs, if further taking a proper transverse momentum dependence, e.g. the BHL-transverse momentum dependence~\cite{bhl}, we can construct a better behaved scalar wave functions in which the end-point singularity can be effectively suppressed. For example, following the similar idea in constructing the pseudo-scalar meson's twist-3 WF model, it is natural to assume that one can obtain reasonable twist-3 contributions to the scalar meson involved high energy processes. \\

\noindent {\bf Acknowledgements: } H.Y. Han thanks Y.M. Wang and Y.J. Sun for helpful discussions. This work was supported in part by the Fundamental Research Funds for the Central Universities under Grant No.CDJXS11100002 and No.CQDXWL-2012-Z002, by Natural Science Foundation of China under Grant No.11075225 and No.11275280, and by the Program for New Century Excellent Talents in University under Grant NO.NCET-10-0882. \\

\appendix

\section{Details for deriving the twist-3 LCDAs}
\label{appdasm}

The LCDAs of the scalar mesons are normalized as $\int_{0}^{1}du\phi_{S}^{s}(u,\mu) =\int_{0}^{1}du\phi_{S}^{\sigma}(u,\mu)=1$. The initial energy scale for the bound state and the non-perturbative matrix elements are set as the values of the Borel parameter $M$. Their values at any other scale $\mu$ can be obtained from the RG equations, for example, the decay constant, the quark mass and the condensates are related by \cite{yang,Hwang}
\begin{eqnarray}
\bar{f}_S(M) &=& \bar{f}_S(\mu) \bigg(\frac{\alpha_s (\mu)}{\alpha_s
(M)}\bigg)^{4/b}, \nonumber\\
m_{q}(M) &=& m_{q}({\mu}) \bigg(\frac{\alpha_s (\mu) }{ \alpha_s (M)}\bigg)^{-4/b}, \nonumber \\
\langle \bar{q} q\rangle_{M} &=& \langle \bar{q} q\rangle_{\mu}
\bigg(\frac{\alpha_s (\mu)}{\alpha_s (M)}\bigg)^{4/b}, \nonumber\\
\langle g_s \bar{q} \sigma   G q\rangle_{M} &=& \langle g_s \bar{q}
\sigma G q\rangle_{\mu} \bigg(\frac{\alpha_s (\mu) }{ \alpha_s
(M)}\bigg)^{-2/3b},  \nonumber\\
\langle \alpha_s G^2 \rangle_{M} &=& \langle \alpha_s G^2 \rangle_{\mu} , \label{RGE}
\end{eqnarray}
where $b= (33 -2 n_f)/3$ with $n_f$ the active flavor number. And the RG equations for Gegenbauer moments are \cite{shifman RG}
\begin{eqnarray}
\langle a_n (\mu)\rangle &=& \langle a_n
(\mu_0)\rangle\bigg(\frac{\alpha_s(\mu_0) }{
\alpha_s(\mu)}\bigg)^{-\gamma^S_n /b}, \nonumber\\
\langle b_n (\mu)\rangle &=& \langle b_n(\mu_0)\rangle\bigg(\frac{\alpha_s(\mu_0)}{
\alpha_s(\mu)}\bigg)^{-\gamma^T_n /b}, \label{momentsRGE}
\end{eqnarray}
where the one-loop anomalous dimensions are
\begin{eqnarray}
\gamma^S_n &=& C_F \bigg(1- \frac{8 }{ (n+1)(n+2)}+4 \sum_{j=2}^{n+1}
\frac{1}{j} \bigg) , \\
\gamma^T_n &=& C_F \bigg(1+4\sum_{j=2}^{n+2} \frac{1}{j}\bigg)
\end{eqnarray}
with $C_F=4/3$.

The moments of the LCDAs for the scalar mesons are calculated under the background field approach. For the purpose, we adopt the following two correlation functions
\begin{widetext}
\begin{eqnarray}
i \int d^4 x e^{i q \cdot x }\langle 0 |T\{\bar{q}_1(x)(i z \cdot
\buildrel\leftrightarrow\over D)^n q_2(x), \bar{q}_2(0) q_1(0)\}|0
\rangle &=& -(z \cdot q)^n I_s^{(n,0)}(q^2),\label{eq:lcdasa} \\
i \int d^4 x e^{i q \cdot x }\langle 0 |T\{\bar{q}_1(x)\sigma_{\mu
\nu}(i z \cdot \buildrel\leftrightarrow\over D)^{n+1} q_2(x),
\bar{q}_2(0) q_1(0)\}|0\rangle &=& i (q_{\mu}z_{\nu}-q_{\nu}z_{\mu})(z
\cdot q)^n I_{\sigma}^{(n,0)} (q^2). \label{eq:lcdasb}
\end{eqnarray}
\end{widetext}

On the one hand, the correlation functions (\ref{eq:lcdasa},\ref{eq:lcdasb}) can be evaluated using operator product expansion in the deep Euclidean region ($-q^2 \gg 0$). With the help of the higher-dimensional mass terms reserved in the expression of the propagators (\ref{eq:propagatorofquark}) and (\ref{eq:propagatorofgluon}), the operator product expansion can be expressed as \begin{widetext}
\begin{eqnarray}
{I_s^{n}(q^2)}_{QCD} &=& \frac{3}{4\pi^2}\int^{1}_0dx(2x-1)^n \bigg \{
[q^2x(x-1)+m_1m_2]\left(1-\ln{\frac{q^2x(x-1)+ m_{12}^2}{\mu^2}}\right)\nonumber \\
&&+(n+2)[q^2x(x-1)+ m_{12}^2]\left(\frac{3}{2}-\ln {\frac{q^2x(x-1)+ m_{12}^2}{\mu^2}}\right)\bigg \} \nonumber  \\
&&+\langle \alpha_s G^2 \rangle\frac{1}{8\pi}\int^1_0dxx(1-x)(2x-1)^n
\bigg \{ -\frac{n+1}{q^2x(x-1)+m_{12}^2}+ \frac{2m_1m_2-m_{12}^2}{[q^2x(x-1)+ m_{12}^2]^2}\bigg \}  \nonumber  \\
&&+\langle g_s^3 f G^3 \rangle \frac{n(n-1)}{96\pi^2}\int^1_0dx(2x-1)^{n-2}
\frac{x^2(1-x)^2}{[q^2x(x-1)+m_{12}^2]^2}\nonumber \\
&&+\bigg \{-\langle \bar{q}_1q_1 \rangle \bigg [ \frac{2m_2+(n+1)m_1}{2(m_2^2-q^2)}+
\frac{[(2n^2 +n)m_1^3 +6nm^2_1 m_2+3m_1 m_2^2]} {6(m_2^2-q^2)^2}+\frac{m_1^2m_2^2(nm_1+m_2)} {(m_2^2-q^2)^3} \nonumber \\
&& +\frac{m_1^3m_2^4}{(m_2^2-q^2)^4}\bigg ] +\langle g_s \bar{q}_1\sigma TGq_1\rangle \bigg [ \frac{[18(n-1)m_2+n(8n-5)m_1]} {36(m_2^2-q^2)^2}+\frac{[(4n-3)m_1+3m_2]m_2^2} {6(m_2^2-q^2)^3} \nonumber \\
&&+\frac{2m_1m_2^4}{3(m_2^2-q^2)^4}\bigg ] - g_s^2\langle \bar{q}_1q_1\rangle^2  \bigg [ \frac{2n^2+7n-12}{81(m_2^2-q^2)^2} +\frac{2m_2^2}{27(m_2^2-q^2)^3}\bigg ] + g^2_s\langle \bar{q}_1q_1\rangle \langle \bar{q}_2q_2\rangle \frac{4}{9} \frac{m_2^2+q^2} {q^2(m_2^2-q^2)^2} \nonumber \\
&&+ g^2_s\langle \bar{q}_1q_1\rangle \langle \bar{q}_2q_2\rangle \frac{2}{9}
\frac{q^2+2m_1m_2}{q^2(m_1^2-q^2)(m_2^2-q^2)} +(-1)^n(q_1 \leftrightarrow q_2, m_1 \leftrightarrow m_2) \bigg \}
\label{eq:opelcdas}
\end{eqnarray}
and
\begin{eqnarray}
{I_{\sigma}^{n}(q^2)}_{QCD}&=&\frac{3(n+1)}{4\pi^2}\int^{1}_0dx(2x-1)^n[q^2x(x-1)+m_{12}^2]
\left(\frac{3}{2}-\rm{ln}\frac{q^2x(x-1)+m_{12}^2}{\mu^2}\right)\nonumber\\
&&-\frac{(n+1)}{24\pi}\langle \alpha_s G^2\rangle \int^1_0 dx(2x-1)^nx(1-x)\bigg \{ \frac{q^2x(x-1)+ m_{12}^2-2m_1m_2}{[q^2x(x-1)+ m_{12}^2]^2} \bigg \} \nonumber\\
&&+\langle g_s^3fG^3\rangle  \frac{(n+1)m_1m_2}{24\pi^2}\int^1_0dx(2x-1)^n
 \frac{x^2(1-x)^2}{[q^2x(x-1)+ m_{12}^2]^3}\nonumber\\
&&+\bigg \{\langle \bar{q}_1q_1\rangle \frac{-m_{1}(n+1)}{6}\bigg [ \frac{3}{m_2^2-q^2}+
\frac{m_1^2(2n+1)}{(m_2^2-q^2)^2}+\frac{2m_1^2m_2^2}{(m_2^2-q^2)^3}\bigg ] \nonumber\\
&&+\langle g_s \bar{q}_1\sigma TGq_1\rangle (n+1) \bigg [ \frac{(8n+1)m_1+6m_2} {36(m_2^2-q^2)^2} +\frac{2m_1m_2^2}{9(m_2^2-q^2)^3} \bigg ] \nonumber\\
&&+g_s^2\langle \bar{q}_1q_1\rangle^2 (n+1)[\frac{-2n+5}{81(m_2^2-q^2)^2} +\frac{4m_2^2}{81(m_2^2-q^2)^3}] +(-1)^n(q_1 \leftrightarrow q_2, m_1 \leftrightarrow m_2) \bigg \} ,
\label{eq:opelcdat}
\end{eqnarray}
\end{widetext}
where $m_{12}^2=m_{1}^{2} x + m_{2}^{2} (1-x)$.

On the other hand, the correlation functions (\ref{eq:lcdasa},\ref{eq:lcdasb})
can be derived by inserting a complete set of quantum states $\Sigma |n\rangle \langle n|$ in the physical region:
\begin{widetext}
\begin{displaymath}
{\mbox{Im} I_s^n(q^2)}_{had}=-\pi \delta(q^2-m_S^2)m_S^2\bar{f}_S^2\langle \xi_s^n \rangle +\frac{3}{4\pi}\int^1_0dx(2x-1)^n\bigg \{(n+3)q^2x(x-1)+(n+2)m_{12}^2
+m_1m_2\bigg \}\theta(q^2-s_s),
\end{displaymath}
\begin{displaymath}
{\mbox{Im} I_{\sigma}^n(q^2)}_{had}=-\pi \delta(q^2-m_S^2)
\frac{n+1}{3}m_S^2\bar{f}_S^2\langle \xi_{\sigma}^n \rangle +\frac{3(n+1)}{4\pi}\int^1_0dx(2x-1)^n\bigg \{q^2x(x-1)+m_{12}^2\bigg \}\theta(q^2-s_{\sigma}).
\end{displaymath}
\end{widetext}
In deriving the above equations, we have implicitly adopted the quark-hadron duality. These two expressions of correlators (\ref{eq:lcdasa},\ref{eq:lcdasb}) can be matched through the dispersion relation
\begin{eqnarray}
\frac{1}{\pi} \int ds \frac{\mbox{Im} I(s)_{had} }
{s-q^2}=I(q^2)_{QCD}.
\end{eqnarray}
One can apply Borel transformation to both sides to suppress the unknown higher dimensional condensates and the continuum contributions as much as possible. For an arbitrary $q^2$ polynomials, the Borel transformation implies~\cite{modern SR}
\begin{eqnarray}
\mathcal{B}_{M^2}=\lim_{\stackrel{-q^2,n \to \infty}{-q^2/n=M^2}}
\frac{(-q^2)^{(n+1)}}{n!}\left( \frac{d}{dq^2}\right)^n,
\end{eqnarray}
where $M$ is the Borel parameter.

Applying the Borel transformation to the dispersion relation, it results in
\begin{displaymath}
\frac{1}{\pi}\int ds \, e^{-s/M^2} \mbox{Im} I(s)_{had}=\mathcal{B}_{M^2} I(q^2)_{QCD} ,
\end{displaymath}
which finally leads to the sum rules (\ref{eq:resultslcdase}, \ref{eq:resultslcdaso}, \ref{eq:resultslcdate}, \ref{eq:resultslcdato}) adopted in the body of the text.

\end{document}